\documentclass[review,12pt,authoryear]{elsarticle}

\usepackage{amssymb}
\usepackage{amsthm}
\usepackage{amsmath}
\usepackage{mathrsfs}
\usepackage{graphicx}
\usepackage{epstopdf}
\usepackage{float}
\usepackage{caption}
\usepackage{subcaption}
\usepackage{bm}
\usepackage{amssymb} %for arrows
\usepackage{extarrows} %for arrows
\usepackage{bbm}
\usepackage{mathrsfs}
\usepackage{cleveref}
\usepackage{soul}
\usepackage{booktabs}
\usepackage{tabu} % To increase the vertical space between cells in tables
\usepackage{longtable}
\usepackage{xcolor} % To use different colors in the fonts
\journal{Journal of the Mechanics and Physics of Solids}

\makeatletter
\def\@author#1{\g@addto@macro\elsauthors{\normalsize%
    \def\baselinestretch{1}%
    \upshape\authorsep#1\unskip\textsuperscript{%
      \ifx\@fnmark\@empty\else\unskip\sep\@fnmark\let\sep=,\fi
      \ifx\@corref\@empty\else\unskip\sep\@corref\let\sep=,\fi
      }%
    \def\authorsep{\unskip,\space}%
    \global\let\@fnmark\@empty
    \global\let\@corref\@empty  %% Added
    \global\let\sep\@empty}%
    \@eadauthor={#1}
}
\makeatother

\begin{document}
\graphicspath{{Fig_paper1/}}

\begin{frontmatter}



\title{A phase field formulation for dissolution-driven stress corrosion cracking}


\author{Chuanjie Cui \fnref{T}}

\author{Rujin Ma\corref{cor1}\fnref{T}}
\ead{rjma@tongji.edu.cn}

%\author{Airong Chen \fnref{T}}

\author{Emilio Mart\'{\i}nez-Pa\~neda\fnref{IC}}
\ead{e.martinez-paneda@imperial.ac.uk}

\address[T]{College of Civil Engineering, Tongji University, 200092, Shanghai, China}

\address[IC]{Department of Civil and Environmental Engineering, Imperial College London, London SW7 2AZ, UK}

\cortext[cor1]{Corresponding author.}
%\cortext[cor2]{Corresponding author.}

\begin{abstract}
We present a new theoretical and numerical framework for modelling mechanically-assisted corrosion in elastic-plastic solids. Both pitting and stress corrosion cracking (SCC) can be captured, as well as the pit-to-crack transition. Localised corrosion is assumed to be dissolution-driven and a formulation grounded upon the film rupture-dissolution-repassivation mechanism is presented to incorporate the influence of film passivation. The model incorporates, for the first time, the role of mechanical straining as the electrochemical driving force, accelerating corrosion kinetics. The computational complexities associated with tracking the evolving metal-electrolyte interface are resolved by making use of a phase field paradigm, enabling an accurate approximation of complex SCC morphologies. The coupled electro-chemo-mechanical formulation is numerically implemented using the finite element method and an implicit time integration scheme; displacements, phase field order parameter and concentration are the primary variables. Five case studies of particular interest are addressed to showcase the predictive capabilities of the model, revealing an excellent agreement with analytical solutions and experimental measurements. By modelling these paradigmatic 2D and 3D boundary value problems we show that our formulation can capture: (i) the transition from activation-controlled corrosion to diffusion-controlled corrosion, (ii) the sensitivity of interface kinetics to mechanical stresses and strains, (iii) the role of film passivation in reducing corrosion rates, and (iv) the dependence of the stability of the passive film to local strain rates. The influence of these factors in driving the shape change of SCC defects, including the pit-to-crack transition, is a natural outcome of the model, laying the foundations for a mechanistic assessment of engineering materials and structures.\\
\end{abstract}

\begin{keyword}

Phase field \sep Finite element method \sep Stress corrosion cracking \sep Passive film \sep Mechanochemistry



\end{keyword}

\end{frontmatter}


\section{Introduction}
\label{Sec:Intro} 

The reliability and structural integrity of materials are generally governed by their interaction with the environment and the multi-physics phenomena resulting from those interactions. Stress corrosion cracking (SCC) is one of these phenomena, arguably the most common yet complex failure mechanism of engineering structures and components \citep{Turnbull2001}. SCC involves the propagation of cracks due to the combined action of mechanical stresses and the environment. It is often facilitated by pre-existing defects, such as corrosion pits, and takes place in a broad range of metallic materials and environments \citep{Raja2011}. However, and despite decades of intensive research, the underlying physical mechanisms governing SCC are still not well understood \citep{Turnbull1993,Crane2016}. A number of mechanistic interpretations have been presented, and models can be classified into two categories. The first one refers to models and mechanisms where crack propagation is driven by the anodic corrosion reaction at the crack tip, including the active path dissolution model \citep{Parkins1996} or models grounded on the film rupture-dissolution-repassivation (FRDR) mechanism \citep{Scully1975,Scully1980,Andresen1988}. The second category includes models where crack growth is cathodic-driven and associated with the ingress and transport of hydrogen, such as the hydrogen-enhanced decohesion (HEDE) mechanism \citep{Oriani1972,Serebrinsky2004,JMPS2020}, hydrogen enhanced localized plasticity (HELP) \citep{Sofronis1995} or adsorption-induced dislocation emission (AIDE) \citep{Lynch1988}. The characteristics of SCC can vary significantly from one material to another, or even between different environments within the same material, hindering a critical evaluation of the different mechanistic interpretations. Moreover, many of these mechanisms are not mutually exclusive and could be acting in concert.\\

The development of predictive models for SCC is not only hindered by the complexity of the physical picture but also by the challenges associated with encapsulating into a numerical framework the different physics of the problem (diffusion, electrochemistry, mechanics). It is particularly challenging from a mathematical and computational perspective to capture how the metal-electrolyte interface evolves over time. SCC failures typically involve the growth of multiple pits, a pit-to-crack transition stage and the propagation of one or multiple cracks \citep{Larrosa2018a}. Capturing the complex morphologies resulting from this three-stage process is not an easy task and typically requires defining moving interfacial boundary conditions and manually adjusting the interface topology with arbitrary criteria when merging or division occurs. Numerical methods have been developed to capture the moving boundary problem of localised corrosion; these include the eXtended Finite Element Method (X-FEM) \citep{Duddu2016}, Arbitrary Lagrangian-Eulerian (ALE) techniques \citep{Sun2014} and peridynamics \citep{Chen2015}. However, progress is often limited by the inability to deal with arbitrary 3D geometries, cumbersome implementations or the computational cost. An alternative approach is the use of the phase field method, which is revolutionising the modelling of moving boundaries and interfacial problems. In the phase field paradigm, the boundary conditions at an interface are replaced by a differential equation for the evolution of an auxiliary (phase) field $\phi$. This phase field takes two distinct values in each of the phases (e.g., 0 and 1), with a smooth change between both values in a diffuse region near the interface. The problem can then be solved by integrating a set of partial differential equations for the whole system, thus avoiding the explicit treatment of the interface conditions. First proposed in the context of solidification and micro-structure evolution, phase field methods are finding ever-expanding applications \citep{Biner2017}, including areas such as fracture mechanics \citep{Bourdin2000,Tanne2018,CPB2019} or fatigue damage \citep{Lo2019,Carrara2020,TAFM2020}. Very recently, phase field models have been developed for environmentally-assisted material degradation, including the propagation of cracks assisted by hydrogen \citep{CMAME2018,Anand2019,Wu2020b} and the evolution of the aqueous electrolyte-metal interface due to material dissolution \citep{Stahle2015,Mai2016,Mai2017,Toshniwal2019,Gao2020}. In addition, recent works have coupled the concepts of phase field evolution due to fracture and material dissolution \citep{Nguyen2017a,Nguyen2017b,Nguyen2018}. \\

In this work, we develop a new phase field formulation for simulating pitting corrosion and stress corrosion cracking. The model builds upon the FRDR mechanism and incorporates the role of mechanics in driving interface kinetics and local film breakage. The predictive capabilities of the model are demonstrated by addressing a number of relevant 2D and 3D problems, including several for which analytical or experimental data exist - a remarkable agreement is observed. The remainder of this manuscript is organised as follows. The theoretical framework presented is described in Section \ref{Sec:Theory}. Details of the finite element implementation are given in Section \ref{Sec:FEM}. In Section \ref{Sec:Results} model predictions are benchmarked against analytical solutions and relevant experimental measurements; practical case studies are also addressed to demonstrate the ability of the model in tackling complex engineering problems. Finally, concluding remarks are given in Section \ref{Sec:Conclusions}.

\section{Theory}
\label{Sec:Theory}

In this Section we formulate our coupled deformation-diffusion-material dissolution theory. First, the key mechanistic assumptions are presented. We proceed then to develop our theory using the principle of virtual work. Subsequently, the governing equations for our theory are formulated, as dictated by the coupled field equations, the free-energy imbalance and a set of thermodynamically-consistent constitutive equations.

\subsection{Phase field FRDR model}

Our formulation aims at capturing the process of film rupture, dissolution and repassivation. Metals and alloys exposed to conditions of passivation are protected by a nm-size impermeable film of metal oxides and hydroxides that effectively isolates the material from the corrosive environment \citep{Macdonald1999}. Independently of the specific cracking mechanism, film-rupture is a necessary condition for localised damage in these conditions. Also, a potential rationale for pitting corrosion and stress corrosion cracking is the rupture of the film due to mechanical straining, followed by localised dissolution of the metal \citep{Scully1980,Jivkov2004}. Repassivation follows each local film rupture event, with the new film being deposited on the bare metal under zero-strain conditions. Further input of mechanical work is thus needed to fracture the new film, in a cyclic process governed by the competition between filming and mechanical straining kinetics. \\

\begin{figure}[H]
\centering
\noindent\makebox[\textwidth]{%
\includegraphics[scale=0.4]{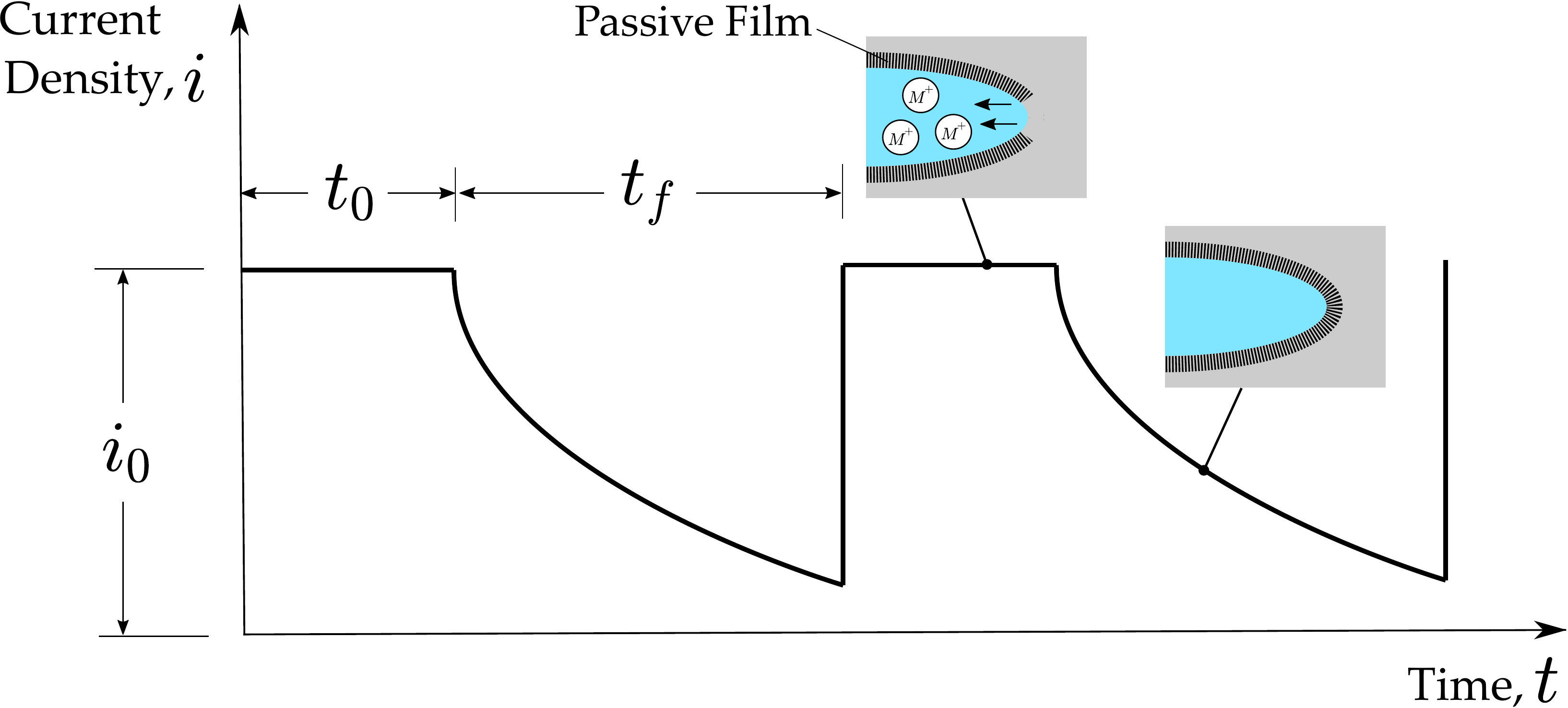}}
\caption{Schematic illustration of two cycles of the film rupture-dissolution-repassivation (FRDR) mechanism at the tip of a defect.}
\label{fig:FRDRmodel}
\end{figure}

As sketched in Fig. \ref{fig:FRDRmodel}, the enhanced corrosion protection resulting from the development of a passive film is characterised by a drop in the corrosion current density $i$, relative to the corrosion current density associated with bare metal $i_0$. After a drop time $t_f$, a film rupture event takes place and the bare metal current density $i_0$ is immediately recovered. This is followed by a time interval $t_0$, during which the current flows before decay begins, such that $t_i=t_0+t_f$ represents one film rupture-dissolution-repassivation cycle. Following common assumptions in the literature \citep{Parkins1987}, the mechanical work required to achieve film rupture can be characterised using an effective plastic strain quantity, $\varepsilon^p$, such that film rupture will take place when the accumulation of the effective plastic strain over a FRDR interval $\varepsilon^p_i$ equals a critical quantity:
\begin{equation}\label{Eq:tf}
    \varepsilon_i^p = \varepsilon_f \,\,\,\,\, \text{with} \,\,\,\,\, \varepsilon_i^p = \int_0^{t_i} \dot{\varepsilon}^p \, \text{d}t
\end{equation}
\noindent where $\dot{\varepsilon}^p$ is the effective plastic strain rate and $\varepsilon_f$ is the critical strain for film rupture, which is on the order of 0.1\% \citep{Gutman2007}. Thus, the duration of each film rupture-dissolution-repassivation cycle, and the magnitude of $t_f$, will be dictated by the local values of the equivalent plastic strain and their evolution in time. We proceed to define the degradation of the current density as an exponential function of the time. Thus, during one rupture-dissolution-repassivation cycle, $i(t_i)$ can be expressed as:
\begin{equation}\label{Eq:cyclei}
    i(t_i)=\left\{
\begin{aligned}
i_0&, \,\,\,\,\,\,\,\,\text{if} \,\,\, \ 0 < t_i \leqslant t_0 \\
i_0 \ \mathrm{exp} \left(-k \left(t_i-t_0 \right)\right)&, \, \,\,\,\,\,\,\,\,\text{if} \,\,\, t_0 < t_i \leqslant t_0+t_f \\
\end{aligned}
\right.
\end{equation}
\noindent where $k$ is a parameter that characterises the sensitivity of the corrosion rates to the stability of the passive film, as dictated by the material and the environment. \\

The role of mechanical stresses and strains is typically restricted to the film rupture event. However, we here consider recent experimental evidence that suggests a further influence of mechanical straining, accelerating corrosion. For example, in \citet{Dai2020} localised corrosion is observed in Q345R steel despite the negligible effect of the passive film in the hydrofluoric acid environment considered. Also, as pointed out by \citet{Gutman2007}, localised corrosion rates can be notably accelerated by the influence of residual stresses. Thus, we enhance the definition of the corrosion current density by considering a mechanochemical term $k_\mathrm{m}$ \citep{Gutman1998} as:
\begin{equation}\label{Eq:Gutman}
    i_a (t) =k_\mathrm{m} \left( \varepsilon^p, \sigma_h \right) i(t_i) = \left(\frac{\varepsilon^p}{\varepsilon_y} + 1 \right) \mathrm{exp}\left(\frac{\sigma_h V_m}{RT}\right) i(t_i) 
\end{equation}

\noindent where $i_a$ is the mechanochemical corrosion current density, $\varepsilon^p$ is the effective plastic strain, $\varepsilon_y$ is the yield strain, $\sigma_h$ is the hydrostatic stress, $V_m$ is the molar volume, $R$ is the gas constant, and $T$ is the absolute temperature. We emphasise that $k_\mathrm{m}$ is a local variable, a function of the local $\varepsilon^p$ and $\sigma_h$ magnitudes.\\

Localised corrosion can be diffusion-controlled or activation-controlled. In the latter, the velocity of the moving pit boundary $\Gamma$ follows Faraday's second law,
\begin{equation}\label{Eq:Faradaylaw}
   v_n=\bm{v} \cdot \bm{\mathrm{n}}=\frac{i_a}{z F c_\mathrm{solid}}
\end{equation}

\noindent where $\bm{\mathrm{n}}$ is the unit normal vector to the pit interface, $F$ is Faraday's constant, $z$ is the average charge number and $c_\mathrm{solid}$ is the concentration of atoms in the metal. Then, the concentration of dissolved ions $c_\mathrm{m}(\bm{\mathrm{x}},t)$ at a point $\bm{\mathrm{x}}$ in the interface can be calculated according to the Rankine–Hugoniot condition as,
\begin{equation}\label{Eq:RH}
   \left[D \nabla c_\mathrm{m} + \left(c_\mathrm{m}(\bm{\mathrm{x}},t)-c_\mathrm{solid} \right)\bm{v}\right] \cdot \bm{\mathrm{n}}=0
\end{equation}

\noindent where $D$ is the diffusion coefficient. 

On the other hand, corrosion becomes diffusion controlled when the surface concentration reaches the saturation concentration $c_\mathrm{sat}$, due to the accumulation of metal ions along the pit boundary. The pit interface velocity becomes then controlled by the diffusion of metal ions away from the pit boundary and the moving pit boundary velocity can be obtained by considering a saturated concentration in (\ref{Eq:RH}), such that
\begin{equation}\label{Eq:v_diffusion}
   v_n=\frac{D \nabla c_\mathrm{m} \cdot \bm{\mathrm{n}}}{c_\mathrm{solid}-c_\mathrm{sat}}
\end{equation}

\begin{figure}[H]
\centering
\noindent\makebox[\textwidth]{%
\includegraphics[scale=0.2]{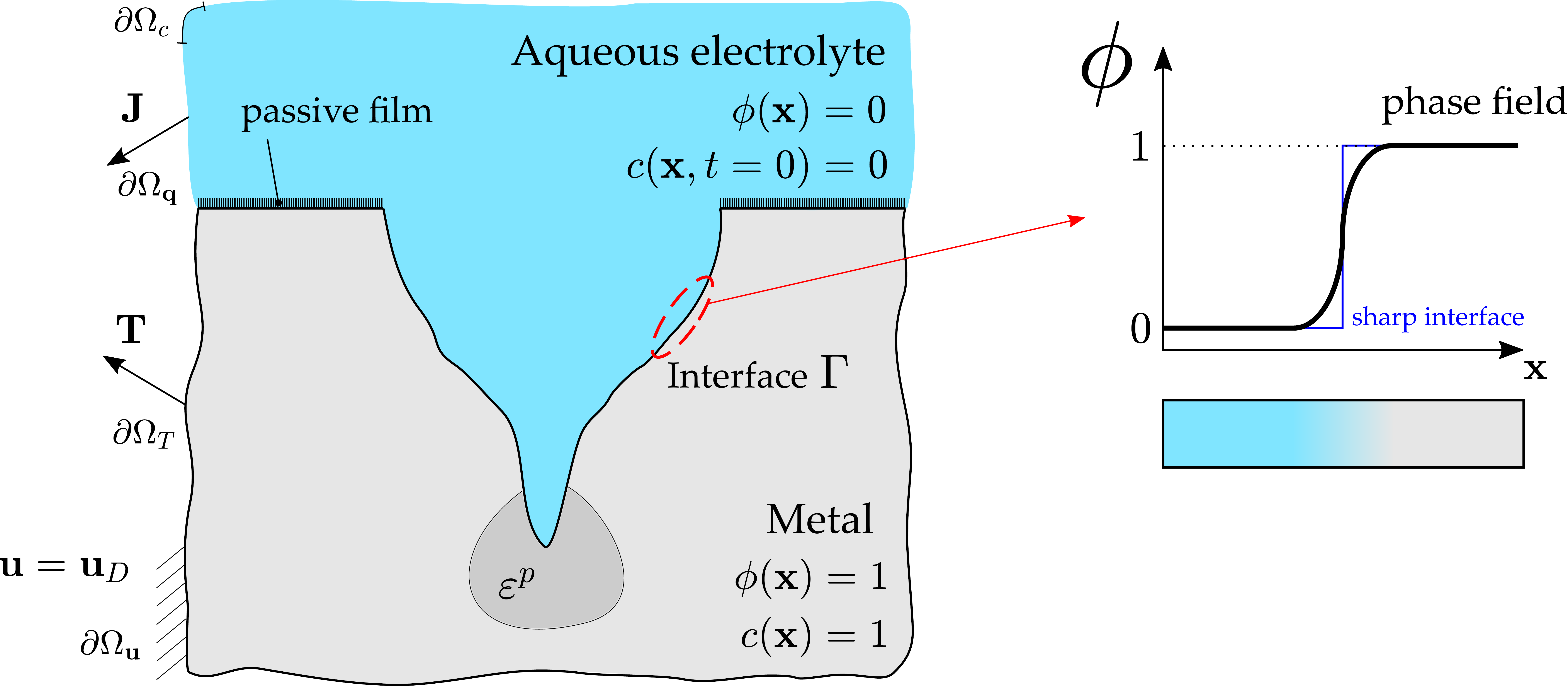}}
\caption{Schematic of the phase field approximation of the localised corrosion damage process, including the electrolyte phase ($\phi=0$), the solid metal phase ($\phi=1$), and the interface $\Gamma$ at a time $t$.}
\label{fig:SCC}
\end{figure}

Modelling the moving pit interface requires the application of both Robin (\ref{Eq:RH}) and Dirichlet ($c_\mathrm{m}=c_\mathrm{sat}$) boundary conditions for the activation-controlled and diffusion-controlled processes, respectively. Instead, we circumvent these complications by following the phase field paradigm proposed by \citet{Mai2016}, approximating the interface evolution implicitly by solving for an auxiliary variable $\phi$. As shown in Fig. \ref{fig:SCC}, the phase field takes values of $\phi=0$ for the electrolyte and $\phi=1$ for the metal, varying smoothly between these two values along the interface $\Gamma$. Also, for consistency, a normalised concentration is defined as $c=c_\mathrm{m}/c_\mathrm{solid}$ \citep{Mai2016}. Thus, the normalised concentration $c$ will equal 1 in the metal (solid) phase, and will approach 0 with increasing distance from the metal-electrolyte interface. \\

\subsection{Principle of virtual work and balance equations}

Our theory deals with the coupled behaviour of two systems, one mechanical and one electrochemical one, see Fig. \ref{fig:SCC}. These two systems are coupled by the following physical phenomena: (i) the pit evolution process is accelerated by mechanics, as presented in Eqs. (\ref{Eq:cyclei}) and (\ref{Eq:Gutman}); and (ii) the dissolution of metals leads to a redistribution of mechanical stress. With respect to the mechanical contributions, the outer surface of the body is decomposed into a part $\partial\Omega_u$, where the displacement can be prescribed by Dirichlet-type boundary conditions, and a part $\partial\Omega_T$, where a traction $\bm{T}$ can be prescribed by Neumann-type boundary conditions. As for the electrochemical part, the external surface consists on two parts: $\partial \Omega_q$ where the flux $\bm{J}$ is known (Neumman-type boundary conditions), and $\partial \Omega_c$, where the concentration is prescribed (Dirichlet-type boundary conditions). Accordingly, a concentration flux entering the body across $\partial \Omega_q$ can be defined as $q=\bm{J} \cdot \bm{n}$. Solute diffusion is driven by the chemical potential $\mu$, as constitutively characterised below. We follow \citet{Duda2018} in defining a scalar field $\eta$ to determine the kinematics of composition changes, such that
\begin{equation}
    \dot{\eta} = \mu \, \, \, \, \, \, \text{and} \, \, \, \, \, \, \eta (\bm{\mathrm{x}}, t) = \int_0^t \mu (\bm{\mathrm{x}}, t) \, \text{d} t 
\end{equation}

Thus, from a kinematic viewpoint, the domain $\Omega$ can be described by the displacement $\bm{u}$, phase parameter $\phi$ and chemical displacement $\eta$. We denote the set of virtual fields as $\left( \delta \bm{u}, \, \delta \phi, \, \delta \eta \right)$ and proceed to introduce the principle of virtual work for the coupled systems as,
\begin{equation}\label{eq:virtualWork}
    \begin{aligned}
    \int_\Omega & \left\{\bm{\sigma} : \delta \bm{\varepsilon} + \frac{1}{L} \frac{d \phi}{d t} \delta \phi + \omega \delta \phi + \bm{\zeta} \cdot \nabla \delta \phi - \frac{d c}{d t} c_\mathrm{solid} \, \delta \eta + \bm{J} \cdot \nabla \delta \eta \right\} \, \text{d} V \\
    &= \int_{\partial \Omega} \left\{ f \delta \phi + q \delta \eta +\bm{T} \cdot \delta \bm{u} \right\} \, \text{d} S
    \end{aligned}
\end{equation}

\noindent where $\bm{\varepsilon}$ is the strain tensor, which is computed from the displacement field in the usual manner $\bm{\varepsilon}=\mathrm{sym} \nabla \bm{u}$ and additively decomposes into an elastic part $\bm{\varepsilon}^e$ and a plastic part $\bm{\varepsilon}^p$. Also, we use $L$ to denote the interface kinetics coefficient, $f$ is the phase field microtraction and $\omega$ and $\bm{\zeta}$ are the microstress quantities work conjugate to the phase field $\phi$ and the phase field gradient $\nabla \phi$, respectively. Operating, and making use of Gauss' divergence theorem:
\begin{equation}\label{eq:PVWb}
    \begin{aligned}
    \int_\Omega & \left\{ \left( \nabla \cdot \bm{\sigma} \right) \cdot \delta \bm{u} + \left( \nabla \cdot \bm{\zeta} - \omega - \frac{1}{L} \frac{d \phi}{d t} \right) \delta \phi + \left( \nabla \cdot \bm{J}  + \frac{d c}{d t} c_\mathrm{solid} \right)  \delta \eta \right\} \, \text{d} V \\
    & = \int_{\partial \Omega} \left\{ \left( \bm{\sigma} \bm{n}  - \bm{T} \right) \cdot \delta \bm{u} + \left( \bm{\zeta} \cdot \bm{n}  - f \right) \delta \phi + \left( \bm{J} \cdot \bm{n}  - q \right) \delta \eta  \right\} \, \text{d} S
    \end{aligned}
\end{equation}

Finally, considering that the left-hand side of Eq. (\ref{eq:PVWb}) must vanish for arbitrary variations, the equilibrium equations in $\Omega$ for each primary kinematic variable are obtained as
\begin{equation}\label{eq:LocalBalance}
    \begin{aligned}
    \nabla \cdot \bm{\sigma} = \bm{0} \\
    \left(\nabla \cdot \bm{\zeta} - \omega \right) -\frac{1}{L} \frac{d \phi}{d t} = 0 \\
    \frac{d c}{d t} c_\mathrm{solid} + \nabla \cdot \bm{J} = 0
    \end{aligned}
\end{equation}

\noindent with the right-hand side of Eq. (\ref{eq:PVWb}) providing the corresponding set of boundary conditions on $\partial \Omega$,
\begin{equation}\label{eq:LocalBalance_BC}
    \begin{aligned}
    \bm{T} = \bm{\sigma} \cdot \bm{n} \\
    f = \bm{\zeta} \cdot \bm{n} \\
    q = \bm{J} \cdot \bm{n}
    \end{aligned}
\end{equation}

\subsection{Energy imbalance}
The first two laws of thermodynamics for a continuum body within a dynamical process of specific internal energy $\mathscr{E}$ and specific entropy $\mit \Lambda$ read \citep{Gurtin2010},
\begin{equation}\label{eq:thermodynamics laws}
    \begin{aligned}
    & \frac{d}{dt} \int_\Omega \mathscr{E} \, \text{d} V = \dot W_e \left( \Omega \right) - \int_{\partial \Omega} \bm{Q} \cdot \bm{n} \, \text{d} S + \int_\Omega Q \, \text{d} V \\
    & \frac{d}{dt} \int_\Omega {\mit \Lambda} \, \text{d} V \geqslant - \int_{\partial \mathrm{\Omega}} \frac{ \bm{Q} }{T} \cdot \bm{n} \, \text{d} S +  \int_\mathrm{\Omega} \frac{Q}{T} \, \text{d} V
    \end{aligned}
\end{equation}

\noindent where $\dot W_e$ is the power of external work, $\bm{Q}$ is the heat ﬂux and $Q$ is the heat absorption. The resulting Clausius-Duhem inequality must be fulfilled by the free energies associated with both systems, mechanical $\psi^M$ and electrochemical $\psi^E$. Thus, the energy imbalances associated with the mechanical and electrochemical systems can be obtained by assuming an isothermal process $(T=T_0)$ and replacing the virtual fields $\left( \delta \bm{u}, \, \delta \phi, \, \delta \eta \right)$ by the realizable velocity fields $\left( \dot{\bm{u}}, \, \dot{\phi}, \, \mu \right)$ in Eq. (\ref{eq:virtualWork}), 
\begin{equation}\label{eq:Energy imbalanceM}
    \frac{d}{dt} \int_\Omega \psi^M \, \text{d} V \leqslant \int_{\partial \Omega}  \bm{T} \cdot \dot{\bm{u}}  \, \text{d} S
\end{equation}
\begin{equation}\label{eq:Energy imbalanceE}
    \frac{d}{dt} \int_\Omega \psi^E \, \text{d} V \leqslant \int_{\partial \Omega} \left(  f \dot{\phi} + q \mu \right) \, \text{d} S
\end{equation}

Employing the divergence theorem and recalling the local balance equations (\ref{eq:LocalBalance})-(\ref{eq:LocalBalance_BC}), we find the equivalent point-wise version of Eqs. (\ref{eq:Energy imbalanceM})-(\ref{eq:Energy imbalanceE}) as,
\begin{equation}\label{eq:Energy imbalance_2M}
    \dot{\psi}^M - \bm{\sigma} :  \dot{\bm{\varepsilon}}  \leqslant 0
\end{equation}
\begin{equation}\label{eq:Energy imbalance_2E}
    \dot{\psi}^E - \frac{1}{L} \ddot{\phi} - \omega \dot{\phi} - \bm{\zeta} \cdot \nabla \dot{\phi}  + \mu \dot{c} \, c_\mathrm{solid} - \bm{J} \cdot \nabla \mu \leqslant 0
\end{equation}

\subsection{Constitutive theory}

We proceed to define the functional form of free energy of each system and subsequently derive a set of thermodynamically consistent relations. 

\subsubsection{Constitutive prescriptions for the mechanical problem}

The mechanical behaviour of the solid is characterised by von Mises J2 plasticity theory. The extension to strain gradient plasticity \citep{EJMAS2019,JMPS2019}, of particular importance when dealing with sharp pits and cracks, will be addressed in future works. Accordingly, the mechanical free energy is decomposed into elastic $\psi^e$ and plastic $\psi^p$ components, both of which are degraded by the phase field,
\begin{equation}
    \psi^M = h (\phi) \left( \psi^e + \psi^p \right)
\end{equation}

\noindent Here, $h(\phi)$ is the degradation function characterising the transition from the undissolved solid ($\phi=1$) to the electrolyte phase ($\phi=0$). The definition of $h(\phi)$ must satisfy the conditions $h(\phi=0)=0$ and $h(\phi=1)=1$; here,
\begin{equation}
    h \left( \phi \right) = -2 \phi^3 + 3 \phi^2
\end{equation}

Consistent with (\ref{eq:Energy imbalance_2M}), the Cauchy stress tensor is defined as $\bm{\sigma}=\partial_{\bm{\varepsilon}} \psi^M$ and we denote $\bm{\sigma}_0$ as the Cauchy stress tensor for the undissolved solid. Thus, accounting for the influence of the phase field and using $\bm{\sigma}_0$, the mechanical force balance (\ref{eq:LocalBalance}a) can be reformulated as,
\begin{equation}\label{eq:Strong_M}
   \nabla \cdot \left[  \left(h\left( \phi \right) + \kappa \right)  \bm{\sigma_0} \right] = \bm{0}
\end{equation}

\noindent where $\kappa$ is a small positive parameter introduced to circumvent the complete degradation of the energy and ensure that the algebraic conditioning number remains well-posed. We adopt $\kappa=1 \times 10^{-7}$ throughout this work.\\

The elastic strain energy density is defined as a function of the elastic strains $\bm{\varepsilon}^e$ and the linear elastic stiffness matrix $\bm{C}_0$ in the usual manner,
\begin{equation}
     \psi^e \left( \bm{\varepsilon}^e \right) = \frac{1}{2} \left(\bm{\varepsilon}^e \right)^T : \bm{C}_0 : \bm{\varepsilon}^e   
\end{equation}

\noindent While the plastic strain energy density is incrementally computed from the plastic strain tensor $\bm{\varepsilon}^p$ and the Cauchy stress tensor for the undissolved solid $\bm{\sigma}_0$ as,
\begin{equation}
\psi^p = \int_0^t \bm{\sigma}_0 : \dot{\bm{\varepsilon}}^p \, \text{d}t
\end{equation}

Finally, the material work hardening is defined assuming an isotropic power law hardening behaviour. Such that the flow stress $\sigma$ and the effective plastic strain $\varepsilon^p$ are related by,
\begin{equation}\label{Eq:plastic}
    \sigma  = \sigma_y \left(1 + \frac{E \varepsilon^p}{\sigma_y} \right)^N
\end{equation}

\noindent where $E$ is Young’s modulus, $\sigma_y$ is the yield stress and $N$ is the strain hardening exponent ($0 \leqslant N \leqslant 1$).

\subsubsection{Constitutive prescriptions for the electrochemical problem}

In the localised corrosion system described in Section \ref{Sec:Intro}, the electrochemical free energy density $\psi^E$ can be decomposed into its chemical and interface counterparts,
\begin{equation}\label{Eq:phie}
    \psi^E = \psi^{ch} + \psi^i 
\end{equation}

The chemical free energy density $\psi^{ch}$ can be further decomposed into the energy associated with material composition and a double-well potential energy \citep{Mai2016}, such that
\begin{equation}\label{Eq:psich}
    \psi^{ch} = h \left( \phi \right) \psi_\mathrm{S}^{ch} + \left( 1- h (\phi) \right) \psi_\mathrm{L}^{ch} + w g \left( \phi \right)
\end{equation}
\noindent where the parameter $w$ is the height of the double well potential $g \left( \phi \right)=\phi^2 (1 - \phi)^2$, and $\psi_\mathrm{S}^{ch}$ and $\psi_\mathrm{L}^{ch}$ denote the chemical free energy density terms associated with the solid and liquid phases, respectively. These latter two terms are defined following the KKS model \citep{Kim1999}, which assumes that each material point is a mixture of both solid and liquid phases with different concentrations but similar chemical potentials. Following this assumption, one reaches the following balances
\begin{equation}\label{Eq:c}
    c = h \left( \phi \right) c_\mathrm{S} + \left[ 1 - h \left( \phi \right) \right] c_\mathrm{L}
\end{equation}
\begin{equation}\label{Eq:dc}
    \frac{\partial \psi_\mathrm{S}^{ch} \left( c_\mathrm{S} \right)}{\partial c_\mathrm{S}} =  \frac{\partial \psi_\mathrm{L}^{ch} \left( c_\mathrm{L} \right)}{\partial c_\mathrm{L}}
\end{equation}

\noindent where $c_\mathrm{S}$ and $c_\mathrm{L}$ are the normalized concentrations of the co-existing solid and liquid phases, respectively. Defining a free energy density curvature $A$, assumed to be similar for the solid and liquid phases, and restricting our attention to dilute solutions, we reach the following definitions for each component of the chemical free energy density,
\begin{equation}\label{Eq:fsfl}
    \psi_\mathrm{S}^{ch} = A (c_\mathrm{S} - c_\mathrm{Se})^2 \,\,\,\,\,\,\,\, \text{and} \,\,\,\,\,\,\,\, \psi_\mathrm{L}^{ch} = A (c_\mathrm{L} - c_\mathrm{Le})^2
\end{equation}

\noindent where $c_\mathrm{Se}=c_\mathrm{solid}/c_\mathrm{solid}=1$ and $c_\mathrm{Le}=c_\mathrm{sat}/c_\mathrm{solid}$ are the normalised \emph{equilibrium} concentrations for the solid and liquid phases. Combining Eqs. (\ref{Eq:psich})-(\ref{Eq:fsfl}) renders,
\begin{equation}\label{Eq:psich2}
    \psi^{ch} = A \left[ c - h (\phi)(c_\mathrm{Se} - c_\mathrm{Le}) - c_\mathrm{Le} \right]^2 + w \phi^2 \left( 1 - \phi \right)^2
\end{equation}

On the other hand, the interface free energy density $\psi_i$ is defined as a function of the gradient of the phase field variable as:
\begin{equation}\label{Eq:phi_i}
    \psi^i = \frac{\alpha}{2} |\nabla \phi|^2 
\end{equation}
\noindent where $\alpha$ is the gradient energy coefficient. In the free energy density definitions (\ref{Eq:psich2})-(\ref{Eq:phi_i}), the parameters $\alpha$ and $w$ govern the interface energy $\gamma$ and its thickness $l$ as:
\begin{equation}\label{eq:gammal}
    \gamma = \frac{\sqrt{\alpha w}}{4}   \,\,\,\,\,\,\,\, \text{and} \,\,\,\,\,\,\,\, l = a^* \sqrt{\frac{2 \alpha }{w}} 
\end{equation}

\noindent where $a^*=2.94$ is a constant parameter corresponding to the definition of the interface region $0.05 < \phi < 0.95$ \citep{Abubakar2015}. The results are sensitive to the choices of $\gamma$ and $l$, which characterise the energy barrier that separates the free energy of the coexisting phases.\\

Finally, the constitutive relations for the associated stress quantities can be readily obtained by fulfilling the free energy imbalance (\ref{eq:Energy imbalance_2E}), which implies
\begin{equation}\label{eq:Energy imbalance_3E}
    - \frac{1}{L} \ddot{\phi}  + \left(\frac{\partial \psi^E}{\partial \phi} - \omega  \right) \dot{\phi} + \left(\frac{\partial \psi^E}{\partial \nabla \phi} - \bm{\zeta} \right) \cdot \nabla \dot{\phi}  + \left(\frac{\partial \psi^E}{\partial c} + \mu c_\mathrm{solid} \right) \dot{c} - \bm{J}  \cdot \nabla \mu \leqslant 0
\end{equation}

Thus, the scalar microstress $\omega$, work conjugate to the phase field $\phi$, is given by
\begin{equation}\label{eq:omega}
    \omega= \frac{\partial \psi^{E}}{\partial \phi} 
    = - 2 A \left[c - h (\phi)(c_\mathrm{Se} - c_\mathrm{Le}) - c_\mathrm{Le} \right] (c_\mathrm{Se} - c_\mathrm{Le}) h ' (\phi) + w g ' (\phi)
\end{equation}

\noindent while the vector microstress $\bm{\zeta}$, work conjugate to the phase field gradient, reads
\begin{equation}\label{eq:zeta}
    \bm{\zeta} = \frac{\partial \psi^E}{\partial \nabla \phi} = \alpha \nabla \phi 
\end{equation}

Inserting these constitutive relations (\ref{eq:omega})-(\ref{eq:zeta}) into the phase field balance (\ref{eq:LocalBalance}b) renders the so-called Allen-Cahn equation:
\begin{equation}\label{eq:Allen–Cahn}
    \frac{d \phi}{d t} + L  \left( \frac{\partial \psi^{E}}{\partial \phi}- \alpha \nabla^2 \phi \right)=0
\end{equation} 

\noindent Eq. (\ref{eq:Allen–Cahn}) shows that material dissolution is governed by the interface kinetics coefficient $L$. The magnitude of $L$ can be considered to be a constant positive number (see, e.g., \citealp{Mai2016}). Here, we assume a time-dependent $L$ instead, enriching the modelling capabilities by establishing a relation with our FRDR mechanistic interpretation and our definition of a mechanochemically-enhanced corrosion current density $i_a (t_i)$. Thus, from Eqs.  (\ref{Eq:cyclei})-(\ref{Eq:Gutman}) and assuming a linear relationship between $L$ and $i_a$, the interface kinetics coefficient over a time interval $t_i$ is defined as
\begin{equation}\label{Eq:cycle}
    L =\left\{
\begin{aligned}
k_\mathrm{m} \left( \varepsilon^p, \sigma_h \right)  L_0 &, \,\,\,\,\,\,\,\,\text{if} \,\,\, 0 < t_i \leqslant t_0 \\
k_\mathrm{m} \left( \varepsilon^p, \sigma_h \right)   L_0 \ \mathrm{exp} \left(-k \left(t_i-t_0 \right)\right) &,  \,\,\,\,\,\,\,\,\text{if} \,\,\, t_0 < t_i \leqslant t_0+t_f \\
\end{aligned}
\right.
\end{equation}

\noindent We emphasise that the interface kinetics coefficient $L$ includes the influence of mechanical straining \textit{via} $k_\mathrm{m} \left( \varepsilon^p, \sigma_h \right)$, and that this is a one-way coupling in that, while $\phi$ reduces the material stiffness, there is no $L$-dependent term in the mechanical problem. Also, we stress that the mechanical free energy density $\psi^M$ does not contribute to the phase field evolution, see (\ref{eq:Allen–Cahn}). In other words, crack advance takes place due to localised metal loss, as postulated by the slip-dissolution model (see, e.g., \citealp{Turnbull1993}). The framework could be extended to incorporate other cracking mechanisms that include a contribution from $\psi^M$; for example, using a coupled framework such as that by \citet{Nguyen2017b}.\\

The material dissolution process (pitting and cracking) can be either activation-controlled or diffusion-controlled, as it will be illustrated in our first case study - Section \ref{eq:CaseStudy1}. The constitutive choices for the mass transport process are obtained as follows. First, the chemical potential is given by,
\begin{equation}\label{eq:mu}
    \mu = - \frac{1}{c_\mathrm{solid}} \frac{\partial \psi^E}{\partial c} =- \frac{2A}{c_\mathrm{solid}}\left(\left(c - h (\phi)(c_\mathrm{Se} - c_\mathrm{Le}) - c_\mathrm{Le} \right)\right)
\end{equation}

\noindent Secondly, the flux $\bm{J}$ can be determined following a Fick law-type relation as,
\begin{equation}\label{eq:flux}
    \bm{J} = \frac{D}{2A} \cdot c_\mathrm{solid} \cdot c_\mathrm{solid} \cdot \nabla \mu = - c_\mathrm{solid} \cdot D \nabla \left(\left(c - h (\phi)(c_\mathrm{Se} - c_\mathrm{Le}) - c_\mathrm{Le} \right)\right)  
\end{equation}

Substituting Eqs. (\ref{eq:mu})-(\ref{eq:flux}) into the mass transport balance Eq. (\ref{eq:LocalBalance}c), the following field equation is obtained
\begin{equation}\label{eq:Cahn–Hilliard}
    \begin{aligned}
     \frac{d c}{d t} - \nabla \cdot D \nabla \left(\left(c - h (\phi)(c_\mathrm{Se} - c_\mathrm{Le}) - c_\mathrm{Le} \right)\right) = 0 
    \end{aligned}
\end{equation}

\noindent which is an extension of Fick's second law such that diffusion of metal ions only takes place along the interface and in the electrolyte.

\section{Numerical implementation}
\label{Sec:FEM}

The finite element method is used to discretise and solve the coupled electro-chemo-mechanical problem. First, we formulate the weak form of the governing equations for the displacement (\ref{eq:Strong_M}), phase field (\ref{eq:Allen–Cahn}), and concentration (\ref{eq:Cahn–Hilliard}) problems, respectively. Thus, 
\begin{flalign}\label{eq:Weak_M}
    \int_\Omega \left(h (\phi)+\kappa \right) \bm{\sigma}_0 \delta \bm{\varepsilon}\,\mathrm{d}V - \int_{\partial \Omega} \bm{T} \cdot \delta \bm{u}\,\mathrm{d}S = 0 &&
\end{flalign} \vspace{-0.8cm}
\begin{flalign}\label{eq:Weak_Allen–Cahn}
    \int_\Omega \frac{\partial \phi}{\partial t} \delta \phi \,\mathrm{d}V + L  \int_\Omega \frac{\partial \psi^{E}}{\partial \phi} \delta \phi \,\mathrm{d}V + L \int_\Omega \alpha \nabla \phi \nabla \delta \phi \,\mathrm{d}V = 0 &&
\end{flalign} \vspace{-0.8cm}
\begin{flalign}\label{eq:Weak_Cahn–Hilliard}
    \int_\Omega \frac{\partial c}{\partial t} \delta c \,\mathrm{d}V + D  \int_\Omega \nabla \left[c - h (\phi)(c_\mathrm{Se} - c_\mathrm{Le}) - c_\mathrm{Le} \right] \cdot \nabla \delta c \, \mathrm{d}V = 0 &&
\end{flalign}

Using Voigt notation, the nodal variables for the displacement field $\bm{\mathrm{\hat{u}}}$, the phase field $\hat{\phi}$ and the normalised concentration $\hat{c}$ are interpolated as:
\begin{equation}
    \bm{u} = \sum_{i=1}^m \bm{\mathrm{N_\emph{i}^u}} \bm{\hat{u}_\emph{i}},\quad \phi = \sum_{i=1}^m N_i \hat{\phi}_i, \quad c = \sum_{i=1}^m N_i \hat{c}_i
\end{equation}

\noindent where $N_i$ denotes the shape function associated with node $i$, for a total number of nodes $m$. Here, $\bm{\mathrm{N_\emph{i}^u}}$ is a diagonal interpolation matrix with the nodal shape functions $N_i$ as components. Similarly, using the standard strain-displacement $\bm{B}$ matrices, the associated gradient quantities are discretised as: 
\begin{equation}
    \bm{\mathrm{\varepsilon}} = \sum_{i=1}^m \bm{\mathrm{B_\emph{i}^u}} \bm{\hat{u}_\emph{i}},\quad \nabla \phi = \sum_{i=1}^m \bm{\mathrm{B_\emph{i}}} \hat{\phi}_i,\quad \nabla c = \sum_{i=1}^m \bm{\mathrm{B_\emph{i}}} \hat{c}_i
\end{equation}

Then, the weak form balances (\ref{eq:Weak_M})-(\ref{eq:Weak_Cahn–Hilliard}) are discretised in time and space, such that the resulting discrete equations of the balances for the displacement, phase field and concentration can be expressed as the following residuals:
\begin{flalign}\label{r_u}
   \bm{r}_{i,\bm{u}}^{n+1} = \int_\Omega \left( h (\phi^{n+1}) + \kappa \right) (\bm{\mathrm{B_\emph{i}^u}})^T \bm{\sigma}_0 \, \mathrm{d}V - \int_{\partial \Omega} (\bm{\mathrm{N_\emph{i}^u}})^T \bm{T} \, \mathrm{d}S  &&
\end{flalign} \vspace{-0.8cm}
\begin{flalign}\label{r_phi}
    r_{i,\phi}^{n+1} = \int_\Omega \frac{\phi^{n+1} - \phi^n}{\mathrm{d\emph{t}}} N_i \,\mathrm{d}V + L \int_\Omega \frac{\partial \psi^{E}}{\partial \phi^{n+1}} N_i \,\mathrm{d}V + L \int_\Omega \alpha \bm{\mathrm{B_\emph{i}}}^T \nabla \phi^{n+1} \,\mathrm{d}V &&
\end{flalign} \vspace{-0.8cm}
\begin{flalign}\label{r_c}
    r_{i,c}^{n+1} = \int_\Omega \frac{c^{n+1} - c^n}{\mathrm{d\emph{t}}} N_i \,\mathrm{d}V 
    + D \int_\Omega \bm{\mathrm{B_\emph{i}}}^T \left[\nabla c^{n+1} - h '  (\phi^{n+1})(c_\mathrm{Se} - c_\mathrm{Le}) \nabla \phi^{n+1} \right] \, \mathrm{d}V &&
\end{flalign}
\noindent where $()^{n+1}$ denotes the $(n+1)$ time step and $\mathrm{d\emph{t}}$ is the time increment. Subsequently, the tangent stiffness matrices are calculated as:
\begin{flalign}\label{K_u}
    \bm{\mathrm{K}}_{ij,\bm{u}}^{n+1} = \int_\Omega \left( h (\phi^{n+1}) + k_p \right) (\bm{\mathrm{B_\emph{i}^u}})^T \bm{C_{ep}} \bm{\mathrm{B_\emph{j}^u}} \, \mathrm{d}V &&
\end{flalign}\vspace{-0.8cm}
\begin{flalign}\label{K_phi}
    \bm{\mathrm{K}}_{ij,\phi}^{n+1} = \int_\Omega \frac{N_i N_j} {\mathrm{d\emph{t}}} \,\mathrm{d}V + L \int_\Omega \frac{\partial ^2  \psi^{E}}{\partial \phi^{2}} N_i N_j\,\mathrm{d}V + L \int_\Omega \alpha \bm{\mathrm{B_\emph{i}}}^T \bm{\mathrm{B_\emph{j}}} \,\mathrm{d}V &&
\end{flalign}\vspace{-0.8cm}
\begin{flalign}\label{K_c}
    \bm{\mathrm{K}}_{ij,c}^{n+1} = \int_\Omega \frac{N_i N_j} {\mathrm{d\emph{t}}} \,\mathrm{d}V + D \int_\Omega \bm{\mathrm{B_\emph{i}}}^T \bm{\mathrm{B_\emph{j}}} \, \mathrm{d}V &&
\end{flalign}

\noindent where $\bm{C_{ep}}$ is the elastic-plastic consistent material Jacobian.

Finally, the linearized finite element system can be expressed as:
\begin{equation}
\left[
\begin{array}{ccc}
    \bm{\mathrm{K}_u} & \bm{0} & \bm{0} \\
    \bm{0} & \bm{\mathrm{K}}_\phi & \bm{0} \\
    \bm{0} & \bm{0} & \bm{\mathrm{K}}_c
\end{array}
\right]
\left[
\begin{array}{c}
    \bm{u} \\ 
    \bm{\phi} \\
    \bm{c}
\end{array}
\right]
=
\left[
\begin{array}{c}
    \bm{r_{u}} \\ 
    \bm{r}_\phi \\
   \bm{r}_c
\end{array}
\right]
\end{equation}

We solve the finite element system by means of a time parametrization and an incremental-iterative scheme in conjunction with the Newton–Raphson method. Good convergence is achieved without the need for staggered schemes; the residual and stiffness matrix components are built using current values and all the equations of the system are solved at once. The numerical model is implemented in the finite element package ABAQUS by means of a user element subroutine (UEL) and Abaqus2Matlab \citep{AES2017} is used for pre-processing the input files. The code, with examples and accompanied documentation, can be downloaded from www.empaneda.com/codes.

\section{Results}
\label{Sec:Results}

We proceed to showcase the potential of the model in predicting localised corrosion with five case studies of particular interest. First, in Section \ref{eq:CaseStudy1}, we validate our predictions with analytical solutions in the context of a pure corrosion problem, the classic benchmark of the pencil electrode test. Secondly, the predictions of the model are compared with experimental measurements of anodic dissolution-driven defect growth in a C-ring test, see Section \ref{Sec:CaseStudy2_Cring}. The influence of mechanical contributions in accelerating and governing the pit-to-crack transition in stainless steel is addressed in Section \ref{Sec:CaseStudy3}. Then, the interaction between multiple pits is investigated in Section \ref{sec:casestudy4} by modelling SCC in a steel plate with pre-existing defects. And finally, the capabilities of the model in conducting large scale 3D simulations of pit corrosion are showcased in Section \ref{eq:casestudy5}.

\subsection{Analytical verification under uniform corrosion: pencil electrode test}
\label{eq:CaseStudy1}

\begin{figure}[H]
\centering
\noindent\makebox[\textwidth]{%
\includegraphics[scale=0.9]{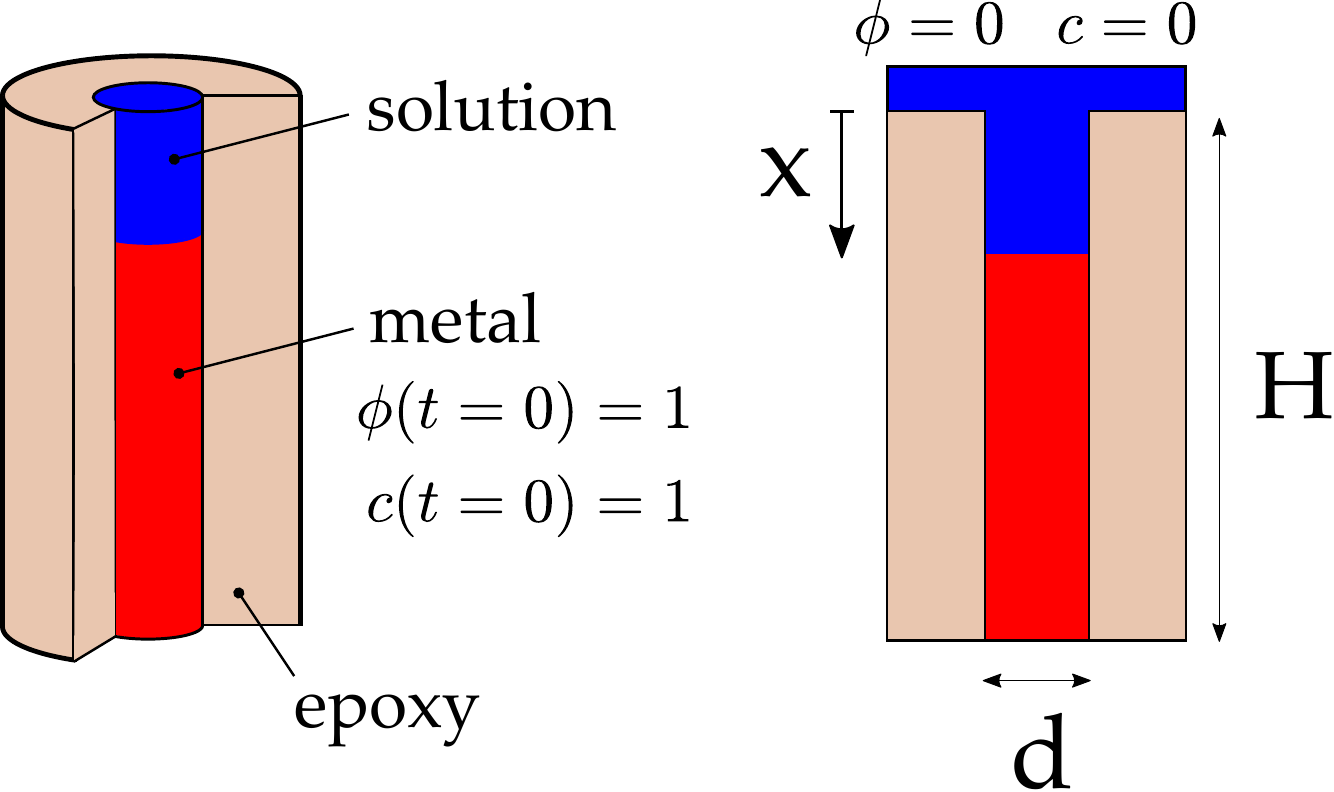}}
\caption{Schematic illustration of the pencil electrode test, including the initial and boundary conditions defined in the phase field model.}
\label{fig:1DpencilSchematic}
\end{figure}

The paradigmatic benchmark of a pencil electrode test is addressed first. This boundary value problem enables validating the numerical framework under pure corrosion conditions, as no mechanical load is applied and no film passivation takes place. The specimen, a H$=$150 $\mu$m long metal wire with diameter d$=25$ $\mu$m, is mounted into an epoxy coating, leaving only one end of the sample exposed to the solution - see Fig. \ref{fig:1DpencilSchematic}. Thus, the problem is essentially one-dimensional and can be described by the change in time of the position of the metal-electrolyte interface $x_\mathrm{d}$. 

The evolution of the pit depth $x_\mathrm{d}$ can be estimated analytically \citep{Scheiner2007}, and is found to be proportional to $\sqrt{t}$ as,
\begin{equation}\label{eq:Scheiner_a}
    x_\mathrm{d} = 2 \xi_\mathrm{d} \sqrt{D t}
\end{equation}

\noindent where the unknown $\xi_\mathrm{d}$ is given by:
\begin{equation}\label{eq:Scheiner_b}
    \frac{c_\mathrm{sat}}{c_\mathrm{solid}-c_\mathrm{sat}} \mathrm{exp} \left(-\xi_\mathrm{d}^2\right) = \sqrt{\pi} \xi_\mathrm{d} \, \mathrm{erf}\left(\xi_\mathrm{d} \right)
\end{equation}

A proportional dependency of the pit depth on $\sqrt{t}$ is also observed experimentally \citep{Ernst2002}. In both the experiments and the analytical study, corrosion is assumed to be diffusion-controlled; this can be achieved by using a sufficiently large applied potential such that reaction rates are high and the surface concentration reaches a saturation magnitude. Using the phase field method, the pencil electrode test can be conveniently modelled without the need of prescribing a moving $c_\mathrm{m}=c_\mathrm{sat}$ Dirichlet boundary condition at the interface \citep{Mai2016}. Thus, as described in Fig. \ref{fig:1DpencilSchematic} we assign as initial conditions $\phi=1$ and $c=1$ in the metallic wire and prescribe the Dirichlet boundary conditions $\phi=0$ and $c=0$ at the pit mouth throughout the numerical experiment. Neumann boundary conditions $\bm{J}=0$ and $f=0$ are considered at the sides, to simulate the insulation provided by the epoxy coating. The interface kinetics coefficient is assumed to be constant $L=L_0$, as there is no mechanical straining or protective film. Following \citet{Mai2016} and \citet{Gao2020}, we consider the material and corrosion parameters shown in Table \ref{tab:corro_para}; the values listed for the interface thickness $l$ and the interface energy $\gamma$ result from the choices of a gradient energy coefficient $\alpha=4.8 \times 10^{-5}$ N and a height of double well potential $w=33.3$ N/mm$^2$, see (\ref{eq:gammal}). A uniform finite element mesh is used, with a characteristic element size 10 times smaller than the interface thickness and employing a total of 15,000 quadratic quadrilateral plane strain elements.\\ 

\begin{table} [!htbp]
  \centering
  \caption{Corrosion and material parameters assumed for the pencil electrode test.}
    \begin{tabular*}{\hsize}{@{}@{\extracolsep{\fill}}lll@{}}
    \toprule
    \textbf{Parameter} & \textbf{Value} & {\textbf{Unit}} \\
    \midrule
    Interface energy $\gamma$ & 0.01  & $\mathrm{N/mm}$ \\
    Interface thickness $l$ & 0.005  & $\mathrm{mm}$ \\
    Diffusion coefficient $D$ & $8.5\times10^{-4}$ & $\mathrm{mm^2/s}$ \\
    Interface kinetics coefficient $L_0$ & $2\times10^6$ & $\mathrm{mm^2/(N \cdot s)}$ \\
    Free energy density curvature $A$ & 53.5 & $\mathrm{N/mm^2}$ \\
    Average concentration of metal $c_\mathrm{solid}$ & 143 & $\mathrm{mol/L}$ \\
    Average saturation concentration $c_\mathrm{sat}$ & 5.1 & $\mathrm{mol/L}$ \\
    \bottomrule
    \end{tabular*}
  \label{tab:corro_para}
\end{table}

\begin{figure}[H]
\centering
\noindent\makebox[\textwidth]{%
\includegraphics[scale=0.25]{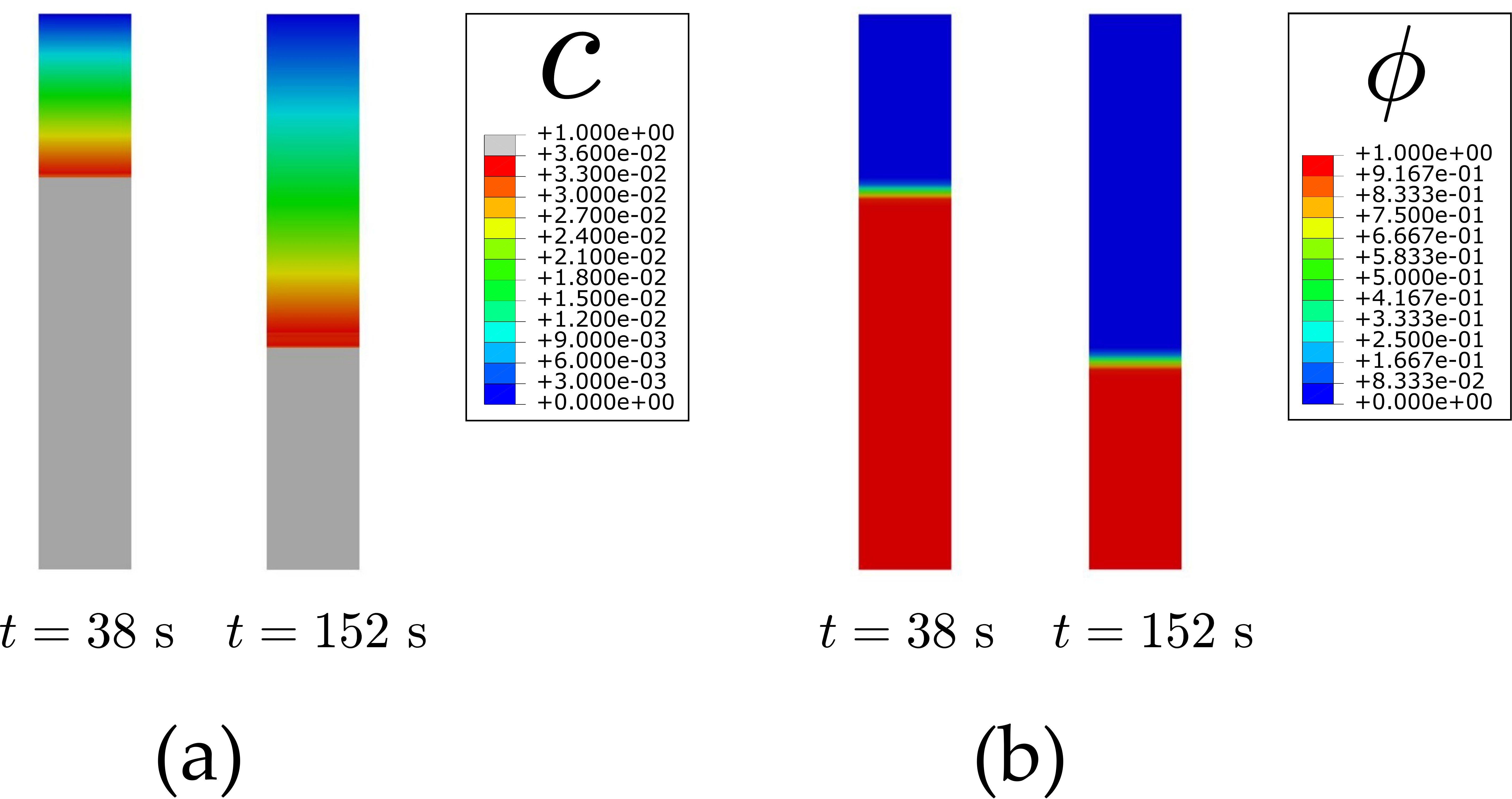}}
\caption{Pencil electrode test. Contours of (a) normalised concentration and (b) phase field at times $t=38\,\mathrm{s}$ and $t=152\,\mathrm{s}$.}
\label{fig:Veri_1_1}
\end{figure}

The contours of the phase field $\phi$ and the normalised concentration $c$ are given in Fig. \ref{fig:Veri_1_1}. Results are shown for two time instants, $38$ and $152$ s, showing how the metal corrodes and the metal-electrolyte evolves. Fig. \ref{fig:Veri_1_1}a shows that the normalised concentration increases from 0 at the pit mouth to $c_\mathrm{sat}/c_\mathrm{solid}=0.036$ at the metal-electrolyte interface; the model captures how a further increase in the pit surface concentration is prevented as saturation occurs. \\

\begin{figure}[H]
\centering
\noindent\makebox[\textwidth]{%
\includegraphics[scale=0.75]{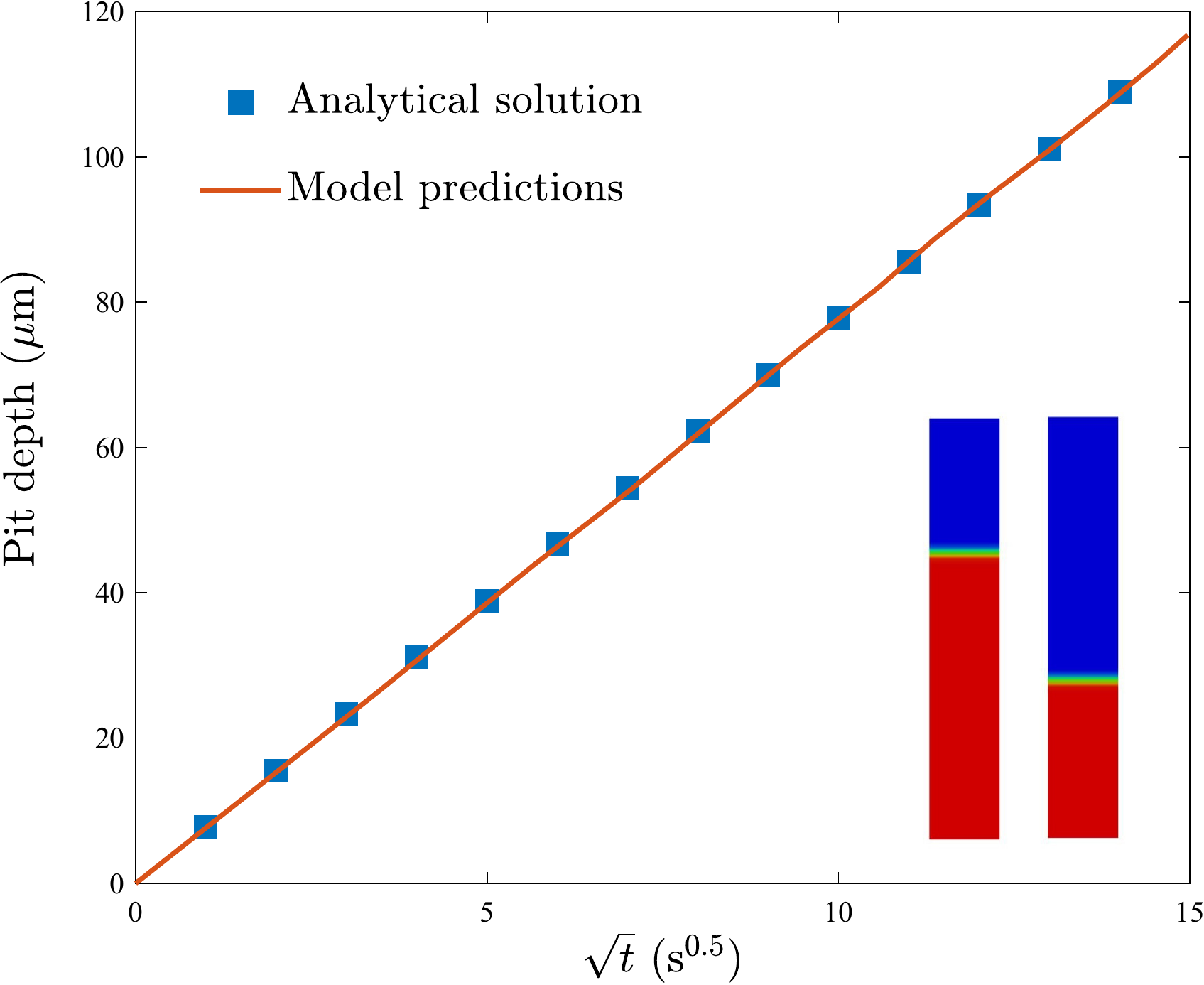}}
\caption{Pencil electrode test. Comparison of model predictions for pit depth versus time against analytical results \citep{Scheiner2007}.}
\label{fig:Veri_1_2}
\end{figure}

The predictions of the model are quantitatively verified against the analytical solution (\ref{eq:Scheiner_a})-(\ref{eq:Scheiner_b}). As shown in Fig. \ref{fig:Veri_1_2}, an excellent agreement is attained in the calculation of the evolution of the pit depth as a function of $\sqrt{t}$. The results obtained also coincide with the numerical predictions by \citet{Mai2016} and \citet{Gao2020}, further validating our numerical implementation.

We proceed to showcase the capabilities of the model in capturing both diffusion-controlled and activation-controlled corrosion. In the pencil electrode test, the corrosion rate is directly controlled by a non-transient parameter $L=L_0$. If $L_0$ is sufficiently high, as in Figs. \ref{fig:Veri_1_1} and \ref{fig:Veri_1_2} (Table \ref{tab:corro_para}), the process is diffusion-controlled, with the electrochemical reaction at the metal-electrolyte interface occurring very fast and leading to saturation in the surface concentration of metal ions. On the other hand, if $L_0$ is small, then the corrosion process is activation-controlled and the pit depth shows a linear dependence with time. For a given time instant, e.g. $t=150$ s, one can readily see that increasing the interface kinetics coefficient raises the magnitude of the pit depth, but this sensitivity to $L_0$ is non-linear and saturates as $L_0$ increases; for a sufficiently large value of $L_0$, further increasing its magnitude has no effect on the corrosion kinetics. Corrosion becomes then diffusion-limited and a higher value of $D$ should be adopted to increase dissolution rates.\\

\begin{figure}[H]
    \centering
    \begin{subfigure}[h]{0.9\linewidth}
    \centering
    \includegraphics[width=1\textwidth]{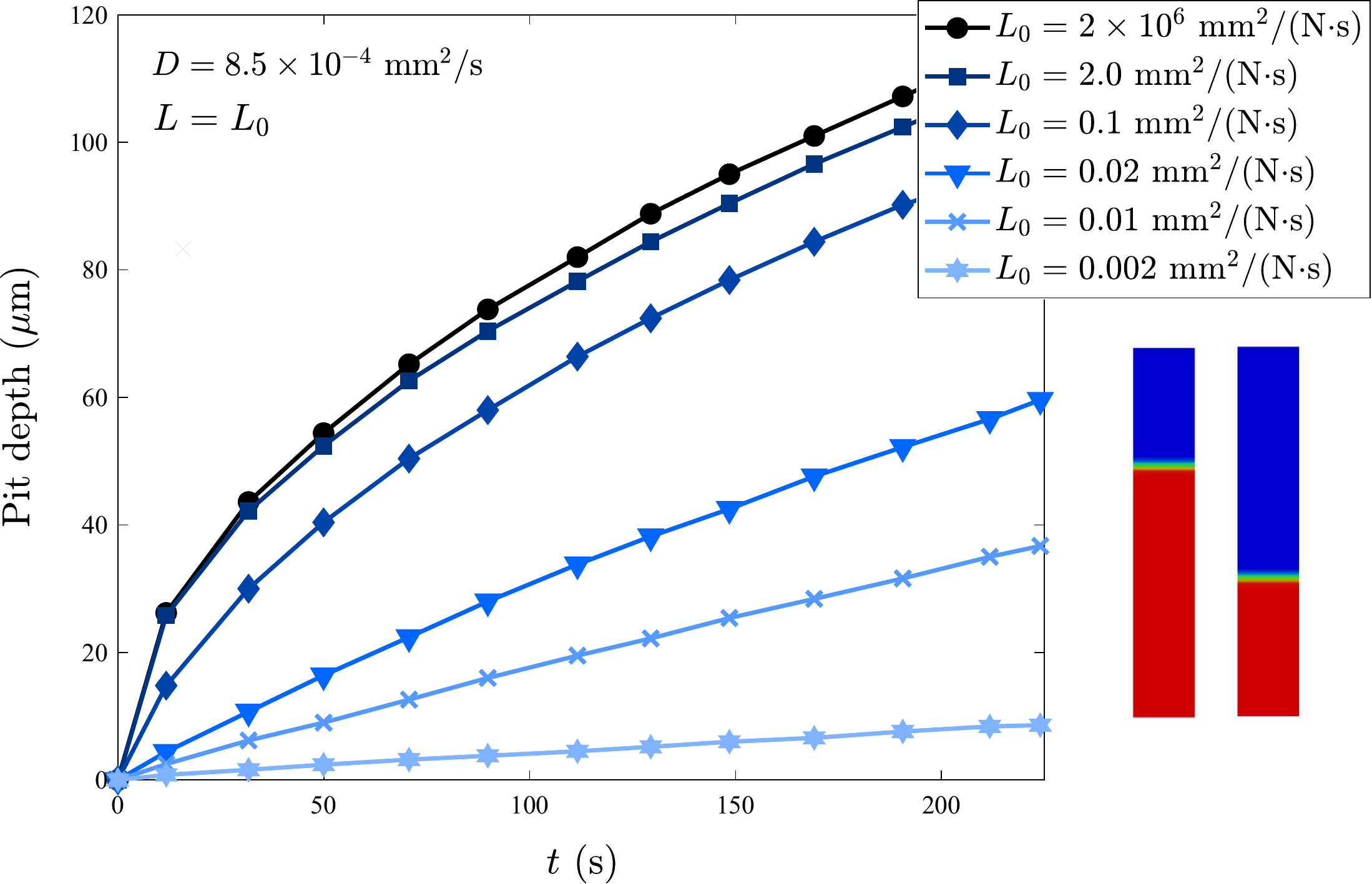}
    \caption{ }
    \label{fig:Veri_1_3}    
    \end{subfigure}
       \begin{subfigure}[h]{0.9\linewidth}
       \centering
    \includegraphics[width=1.05\textwidth]{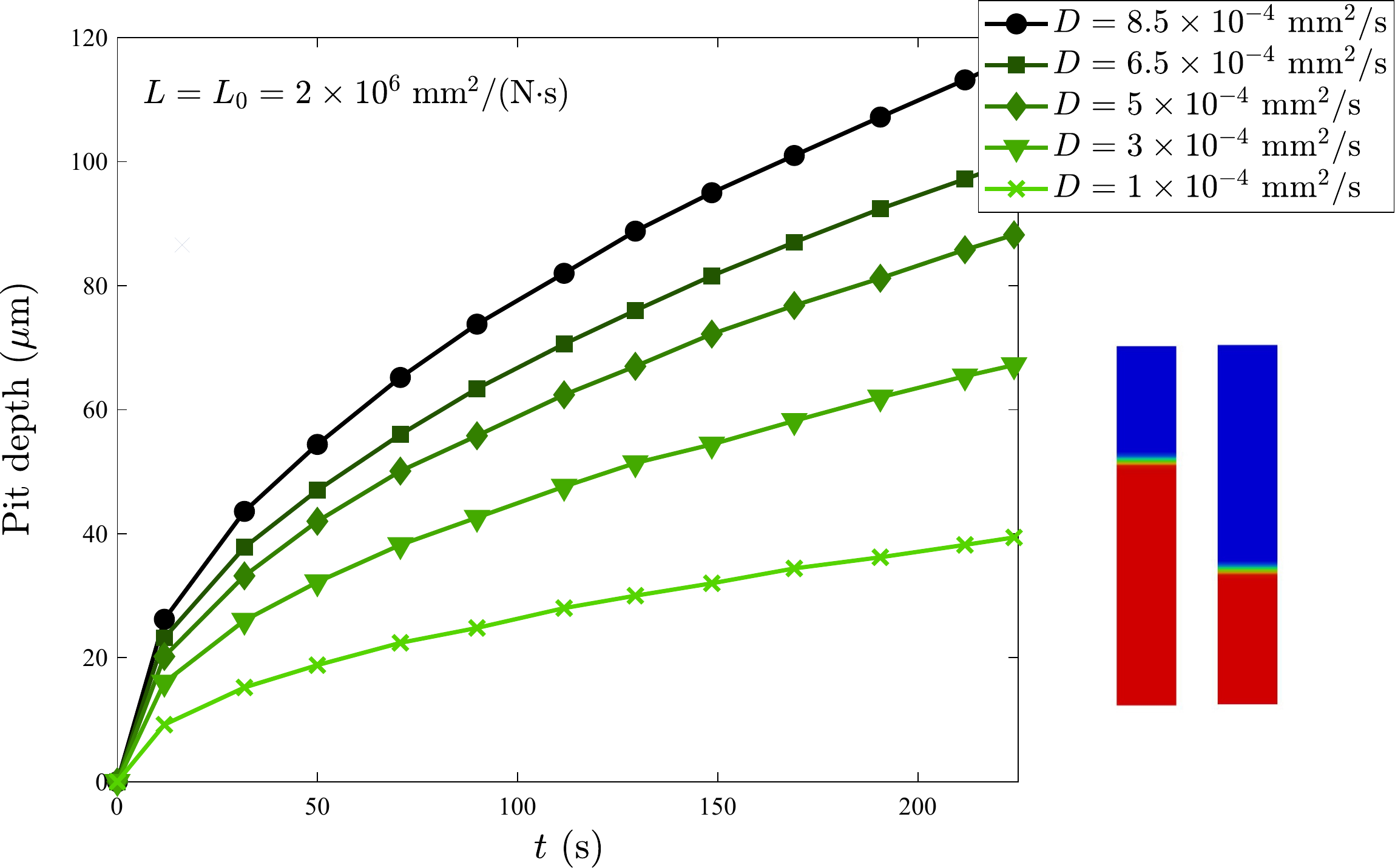}
    \caption{}
    \label{fig:Veri_1_4}    
    \end{subfigure} 
\caption{Pencil electrode test. Pit depth versus time $t$ as a function of (a) the interface kinetics coefficient $L=L_0$, and (b) the diffusion coefficient $D$.}
\end{figure}

Model predictions of pit growth versus time as a function of the interface kinetics parameter are shown in Fig. \ref{fig:Veri_1_3}, exhibiting a linear to parabolic trend with increasing $L_0$. The pit depth evolution is also computed for a large interface kinetics parameter, $L_0=2\times10^6 \, \mathrm{mm^2/(N \cdot s)}$, and selected values of the diffusion coefficient $D$, see Fig. \ref{fig:Veri_1_4}. As expected, the corrosion reaction exhibits a significant sensitivity to the magnitude of the diffusion coefficient. As the influence of mechanical straining and film re-passivation and rupture are incorporated, the magnitude of $L$ will evolve in time and the transition from activation-controlled to diffusion-controlled corrosion will be captured by the model.

\subsection{Chemo-mechanical validation with experiments: C-ring test on Q345R steel}
\label{Sec:CaseStudy2_Cring}

We now verify our model by reproducing the experimental observations by \citet{Dai2020} on C-ring specimens. Q345R steel samples were exposed to a hydrofluoric acid corrosion environment and the growth of pits and associated cracks was quantified for different levels of applied mechanical stress. This choice of environment translates into a negligible influence of film passivation, as the corrosion products are mainly fluorides (not oxides) and the film is not dense enough to prevent the attack of aggressive ions \citep{Dai2020}. The applied strain is held constant throughout the experiment and thus the sensitivity of defect growth rates to mechanical stresses can be quantified. This boundary value problem enables us to validate the chemo-mechanical coupling introduced in our formulation to characterise the interplay between mechanics and corrosion kinetics, see Eq. (\ref{Eq:Gutman}). Note that the sensitivity of corrosion rates to applied stresses cannot be captured with conventional SCC models that restrict the influence of mechanical straining to the local film breakage mechanism (see, e.g., \citealp{Mai2017}).\\

\begin{figure}[H]
\centering
\noindent\makebox[\textwidth]{%
\includegraphics[scale=0.75]{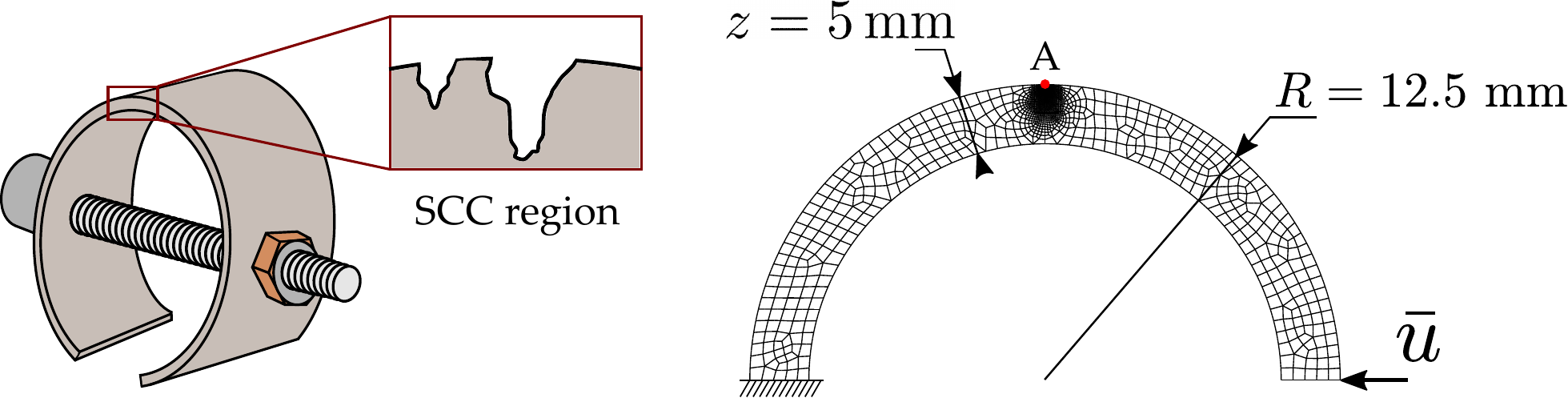}}
\caption{Schematic description of the C-ring SCC test including the geometry, finite element mesh, boundary conditions and dimensions.}
\label{fig:Veri_2_1}
\end{figure}

The configuration of the experiment is shown in Fig. \ref{fig:Veri_2_1}. The sample is subjected to a constant strain by tightening a bolt centred in the diameter of the ring. The load level is characterised by the circumferential stress at the top of the sample, the region where localised corrosion will develop. Thus, the magnitude of the circumferential stress $\sigma_c$ is determined by placing a strain gauge in the middle of the arc, close to the extreme edge; i.e., in the vicinity of point A, see Fig. \ref{fig:Veri_2_1}. The C-ring specimens were exposed to 40 wt.\% hydrofluoric acid under room temperature conditions and, using 3D optical microscopy, surface morphologies were characterised as a function of the stress level and the time \citep{Dai2020}.\\

In the finite element model we discretise the domain between the two bolt holes. As shown in Fig. \ref{fig:Veri_2_1}, the sample has an inner diameter of 20 mm and an outer diameter of 25 mm. The vertical and horizontal displacements are constrained at the left side, while a Dirichlet boundary condition is used to prescribe the horizontal displacement component of the right side, where the bolt is tightened. The load is prescribed at the beginning of the analysis and then held fixed throughout the calculation. The resulting circumferential stress is measured at the top of the sample (point A), to resemble the experimental measurements. In agreement with the experimental work, we consider a Young's modulus of $E=200$ GPa, a Poisson's ratio of $\nu=0.3$ and a yield stress of $\sigma_y=535$ MPa for the Q345R steel. No stress-strain curve or strain hardening exponent is provided in \citep{Dai2020}, a value of $N=0.073$ is adopted based on the Q345R material characterisation by \citet{Liang2019}. The material properties are listed in Table \ref{tab:mecha_para}.

\begin{table} [!htbp]
  \centering
  \caption{Material properties for the Q345R steel used in the C-ring experiments.}
    \begin{tabular*}{\hsize}{@{}@{\extracolsep{\fill}}lll@{}}
    \toprule
    \textbf{Parameter} & \textbf{Value} & {\textbf{Unit}} \\
    \midrule
    Young’s modulus $E$ & 200,000  & MPa \\
    Poisson’s ratio $\nu$ & 0.25  & --- \\
    Yield stress $\sigma_y$ & $535$ & MPa \\
    Strain hardening exponent $N$ & $0.073$ & --- \\
    \bottomrule
    \end{tabular*}
  \label{tab:mecha_para}
\end{table}

\begin{figure}[H]
\centering
\noindent\makebox[\textwidth]{%
\includegraphics[scale=0.28]{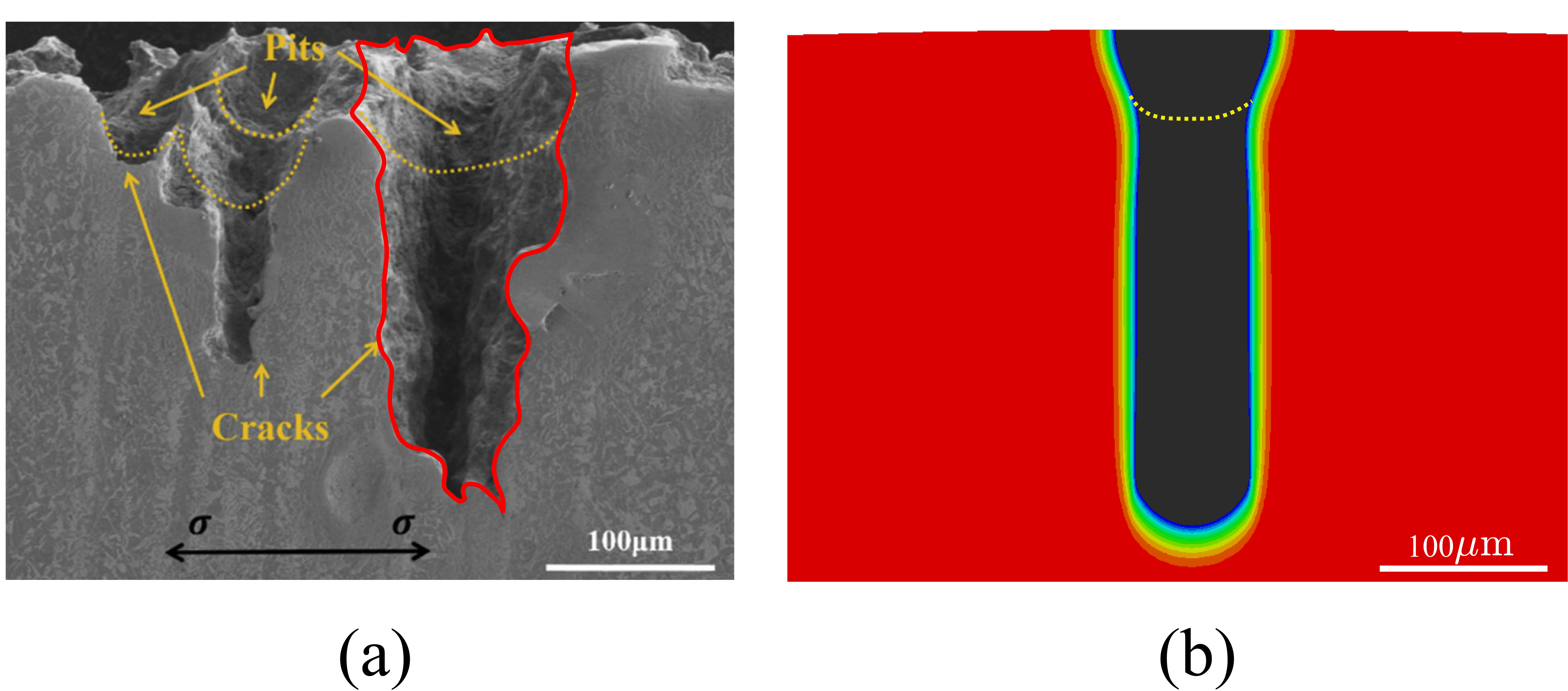}}
\caption{C-ring test, SCC morphology for a stress level $\sigma_c = 0.8 \, \sigma_y$ after $t=210 \, \mathrm{min}$: (a) observed by \citet{Dai2020}, where the red contour highlights the dominant defect that we aim at reproducing; and (b) simulated by our model, where the SCC morphology contour is shown in black.}
\label{fig:Veri_2_2}
\end{figure}

In regard to the corrosion parameters, those listed in Table \ref{tab:corro_para} are used unless otherwise stated. Since no film passivation is involved, we define $k=0$ in Eq. (\ref{Eq:cycle}). The initial interface kinetics coefficient $L_0$ is chosen so as to provide the best fit to the experimental data; here, $L_0=4.7\times10^{-5}\, \mathrm{mm^2/(N \cdot s)}$. The interface thickness is taken to be $l=0.05$ mm and the finite element mesh is refined in the SCC region accordingly, with the characteristic element size being equal or smaller than 0.005 mm. A total of 10,681 quadrilateral quadratic elements are used. As shown in Fig. \ref{fig:Veri_2_2}, several pits are observed in the experiments, with one of them becoming a particularly large crack - we aim at capturing this dominant localised corrosion defect and address later the occurrence of multiple pits and their interactions. To trigger the dominant localised defect, the Dirichlet boundary condition $\phi=0$ is prescribed over a length of 0.05 mm in the extreme edge of the middle of the arc (point A). The finite element results obtained in terms of SCC defect length and shape are provided in Fig. \ref{fig:Veri_2_2}; the experimental measurements from \citet{Dai2020} corresponding to the same time ($t=$210 min) and stress level ($\sigma_c=0.8\sigma_y$) are also shown.\\

\begin{figure}[H]
\centering
\noindent\makebox[\textwidth]{%
\includegraphics[scale=0.8]{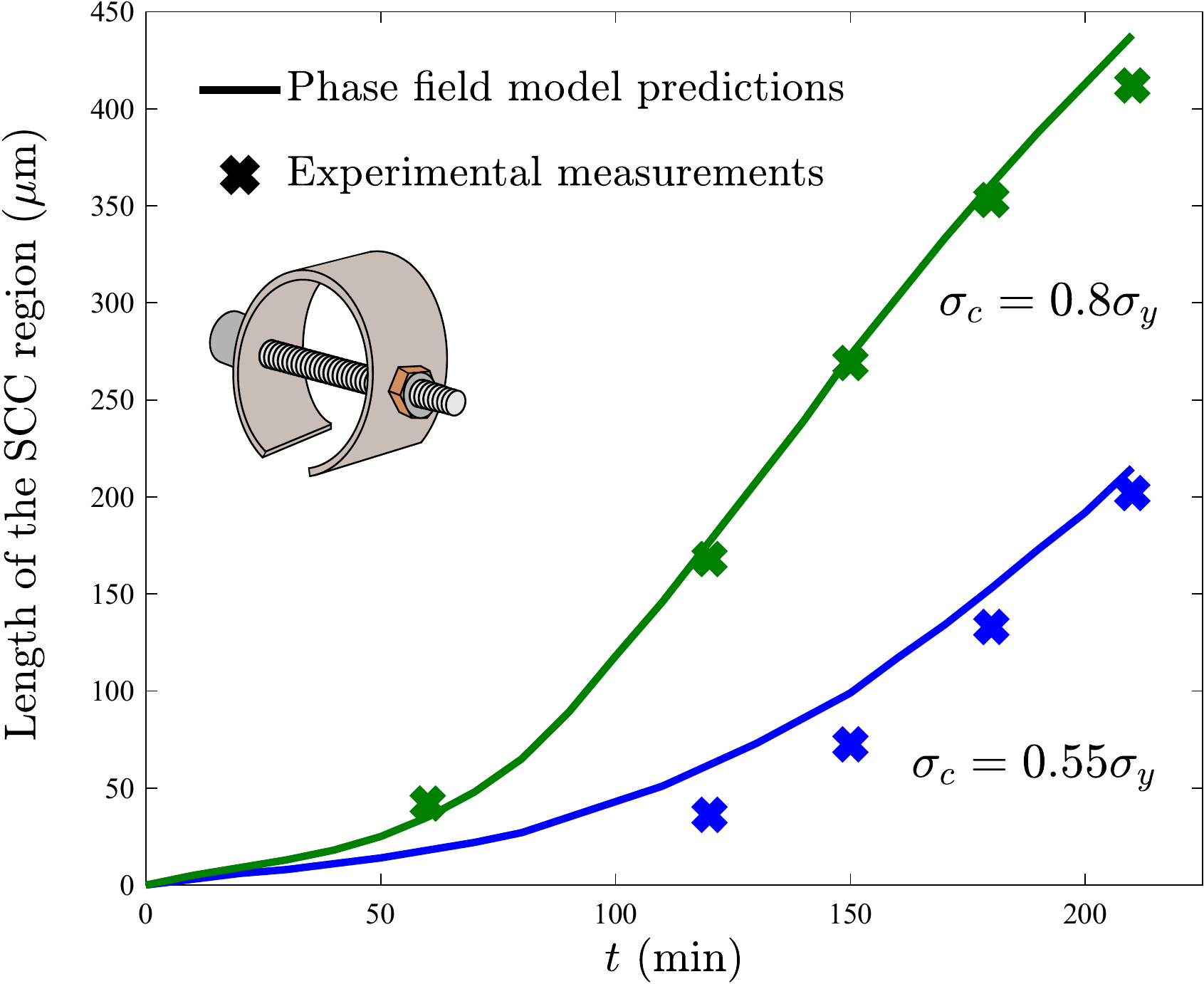}}
\caption{C-ring test. Length of the SCC region versus corrosion time $t$ for two different mechanical stress levels.}
\label{fig:Veri_2_3}
\end{figure}

Fig. \ref{fig:Veri_2_2} reveals a good agreement with experimental observations. A quantitative comparison with experiments for different stress levels is provided in Fig. \ref{fig:Veri_2_3}. Results are shown in terms of the length of the SCC region, including the initial pitting and subsequent cracking, as a function of time. A very good agreement is observed for both $\sigma_c=0.55\sigma_y$ and $\sigma_c=0.8\sigma_y$. We emphasise that while the circumferential stress in the middle of the arc remains below the yield stress, plasticity can develop on the surface of the SCC defects as the resulting notches and cracks act as stress concentrators. 

\subsection{Parametric study: cracking from a semi-elliptical pit}
\label{Sec:CaseStudy3}

\begin{figure}[H]
\centering
\noindent\makebox[\textwidth]{%
\includegraphics[scale=0.8]{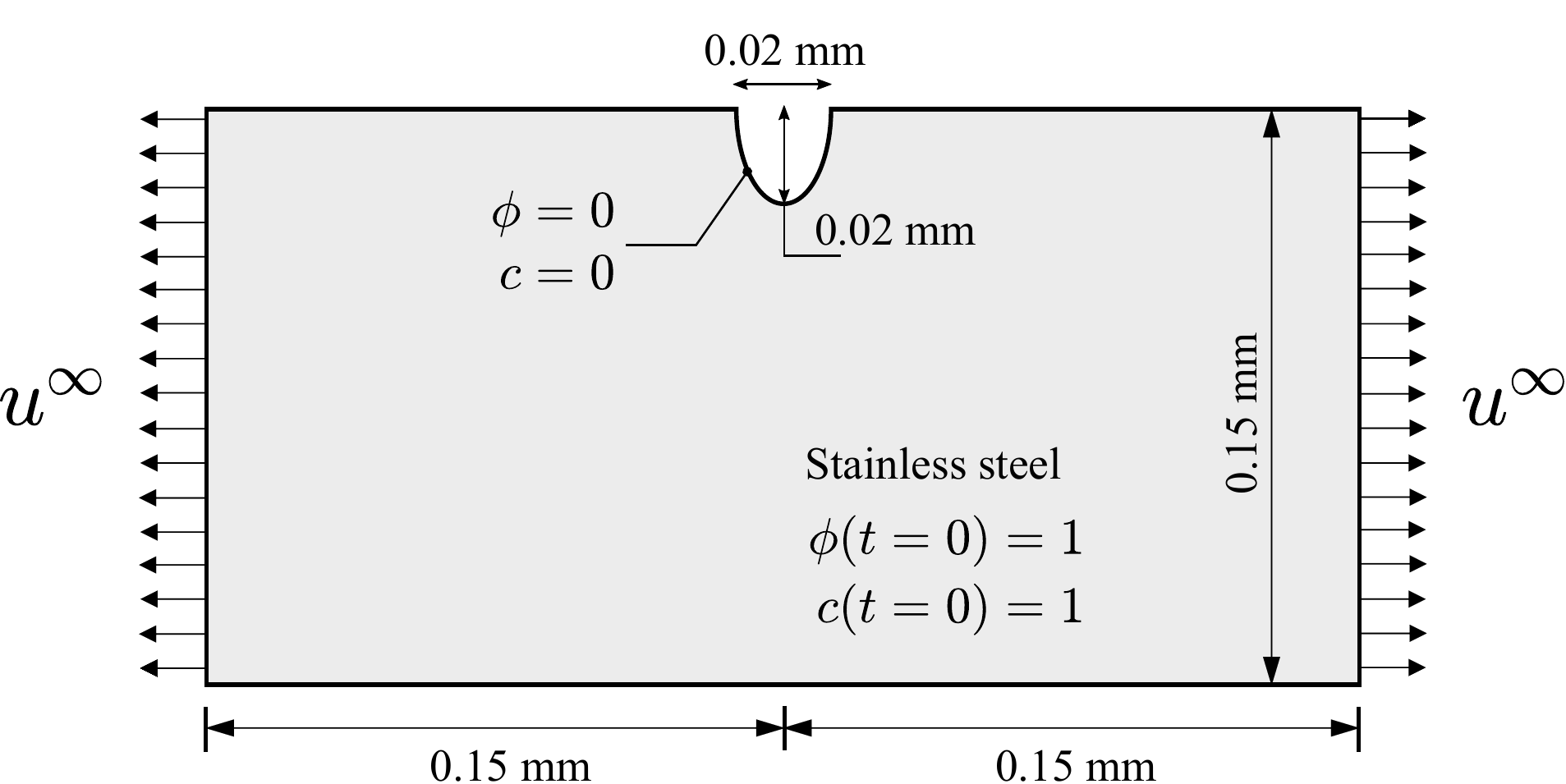}}
\caption{SCC from a semi-elliptical pit: geometric setup, with the initial and boundary conditions.}
\label{fig:Veri_3_1}
\end{figure}

We proceed to conduct numerical experiments on a stainless steel plate, containing a pre-existing pit, to investigate the effect of both mechanical straining and repassivation on the SCC process. The geometric setup, dimensions and boundary conditions are given in Fig. \ref{fig:Veri_3_1}. The Dirichlet boundary conditions $\phi=0$ and $c=0$ are prescribed at the surface of the pre-existing defect, which has a semi-elliptical shape with radii 0.01 and 0.02 mm. At $t=0$, $c=1$ and $\phi=1$ are defined as initial conditions in the metallic plate, of height 0.15 mm and width of 0.3 mm. The mechanical load is applied by prescribing a remote horizontal displacement $u^\infty$ on the left and right edges of the sample, as shown in Fig. \ref{fig:Veri_3_1}. The vertical displacement of the top right corner is also constrained to prevent rigid body motion. 

The corrosion parameters in this study are those assumed in the pencil electrode test, see Table \ref{tab:corro_para}, with the exception of the interface kinetics coefficient; a magnitude of $L_0=0.001 \, \mathrm{mm^2/(N \cdot s)}$ is adopted to ensure activation-controlled corrosion. The constitutive behaviour of the steel plate is characterised by a Young modulus of $E=190$ GPa, a Poisson's ratio of $\nu=0.3$, an initial yield stress of $\sigma_y=520$ MPa and a strain hardening exponent of $N=0.067$. Approximately 7,000 quadrilateral quadratic plane strain elements are used to discretise the model.\\

\begin{figure}[H]
\centering
\noindent\makebox[\textwidth]{%
\includegraphics[scale=0.2]{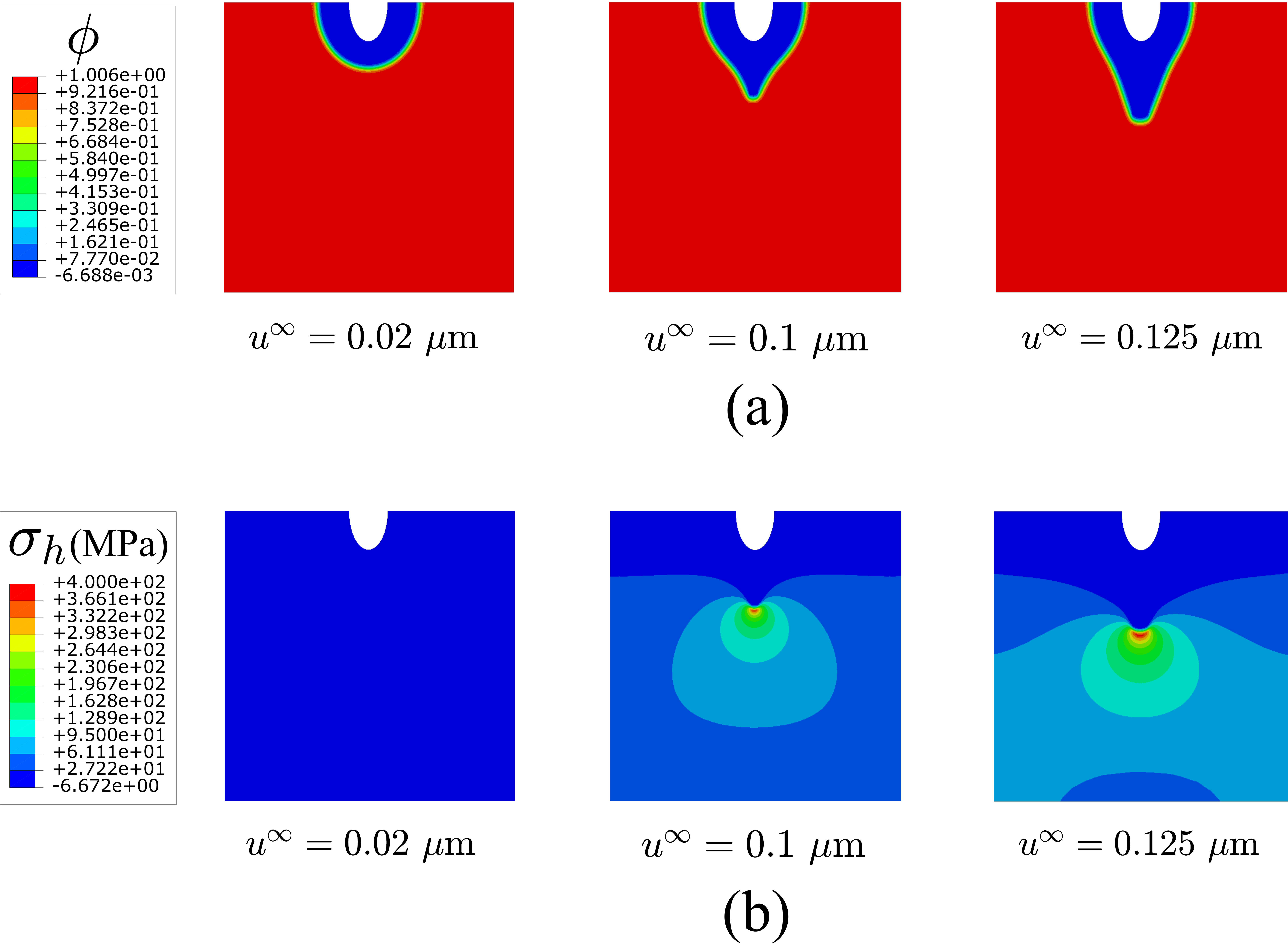}}
\caption{SCC from a semi-elliptical pit. Contours of (a) the phase field and (b) hydrostatic stress for three selected values of the initial applied displacement after $t=900$ s.}
\label{fig:Veri_3_2}
\end{figure}

First, we consider the case of mechanics-enhanced corrosion in the absence of a passivation mechanism. The goal is to characterise the role of mechanical straining in controlling the defect growth rate and shape. Thus, we make $k=0$ in Eq. (\ref{Eq:cycle}) and prescribe three different values of the remote displacement $u^\infty=0.02$ $\mu$m, $u^\infty=0.1$ $\mu$m and $u^\infty=0.125$ $\mu$m. The remote displacement $u^\infty$ is prescribed at the beginning of the numerical experiment and then held fixed over a total time of 900 s. The results obtained, in terms of the phase field and hydrostatic stress contours, are shown in Fig. \ref{fig:Veri_3_2}. 

\begin{figure}[H]
\centering
\noindent\makebox[\textwidth]{%
\includegraphics[scale=0.9]{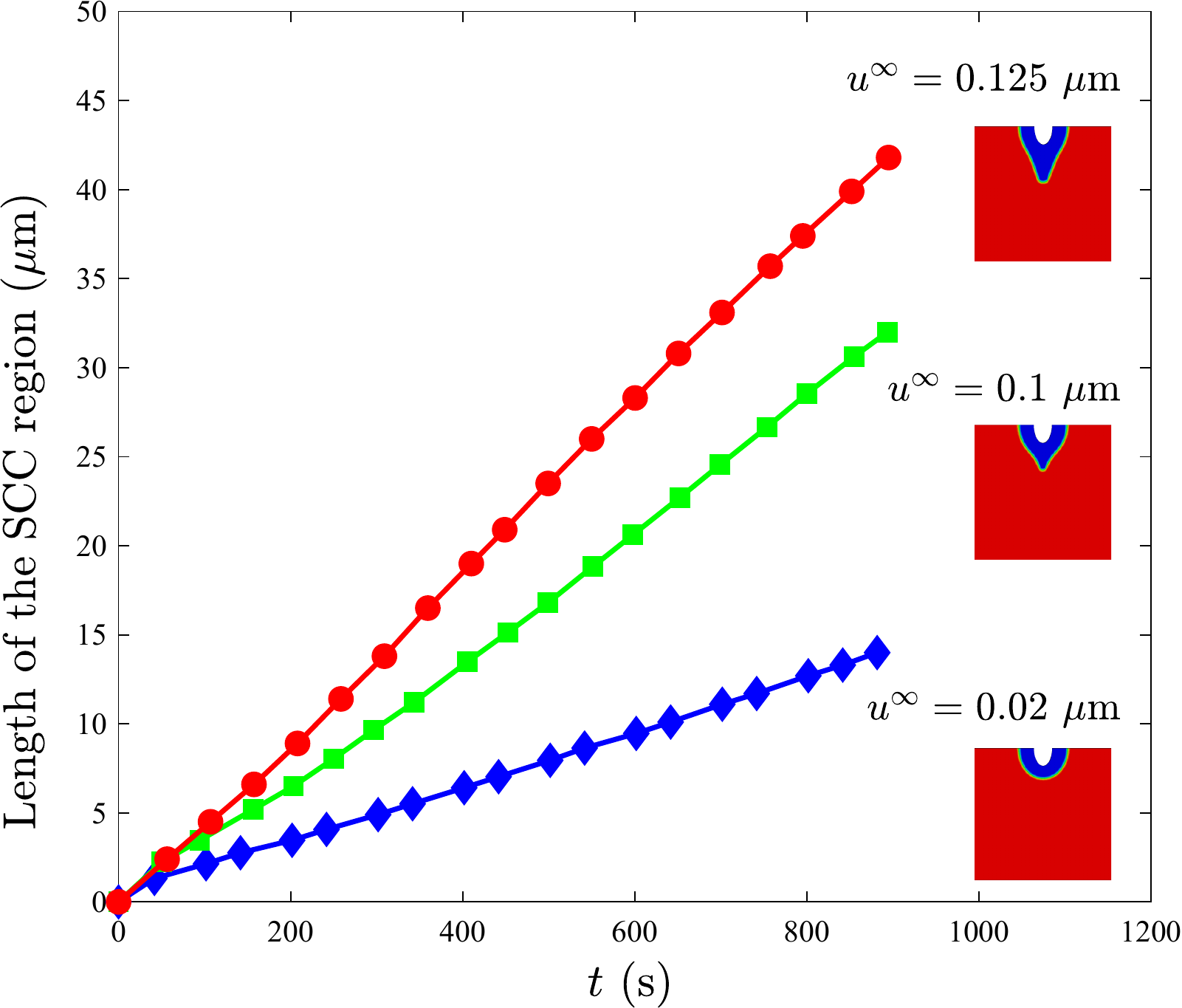}}
\caption{SCC from a semi-elliptical pit. Evolution in time of the SCC region for three different mechanical loads, as characterised by the remote displacement $u^\infty$.}
\label{fig:Veri_3_4}
\end{figure}

The results reveal that, when the mechanical load is small ($u^\infty=0.02$ $\mu$m), there is a negligible influence of the hydrostatic stress and the pit grows approximately the same length in all dimensions, in agreement with the corrosion evolution laws in the absence of mechanical contributions. However as the remote displacement is raised and the magnitude of $\sigma_h$ increases, the pit morphology changes and a pit-to-crack transition is observed. The dominance of the mechanics contributions becomes very clear in the case of $u^\infty=0.125$ $\mu$m, where the stresses are sufficiently high to trigger plastic deformations and the shape of the SCC defect is governed by both the hydrostatic stress and the plastic strain distribution. It is also worth emphasising that mechanical straining not only sharpens the SCC defect but also increases its length. The length of the SCC region is shown in Fig. \ref{fig:Veri_3_4} for the three chosen values of the applied displacement $u^\infty$. It can be seen that the mechanical contribution has a significant effect on localised corrosion rates.

We now turn our attention to the repassivation process and our film rupture-dissolution-repassivation (FRDR) formulation. Thus, we apply and hold fixed a remote displacement of $u^\infty=0.1$ $\mu$m and consider the sensitivity of the results to changes in the material parameter characterising the corrosion sensitivity to the stability of the passive film, $k$ - see Eq. (\ref{Eq:cyclei}) and (\ref{Eq:cycle}). Three representative values are considered: $k=2 \times 10^{-4}$, $k=5 \times 10^{-4}$, and $k=1 \times 10^{-3}$; while the critical strain for film rupture and the time interval without repassivation are held constant at $\varepsilon_f=3 \times 10^{-3}$ and $t_0= 10$ s, respectively.\\ 

\begin{figure}[H]
\centering
\noindent\makebox[\textwidth]{%
\includegraphics[scale=0.2]{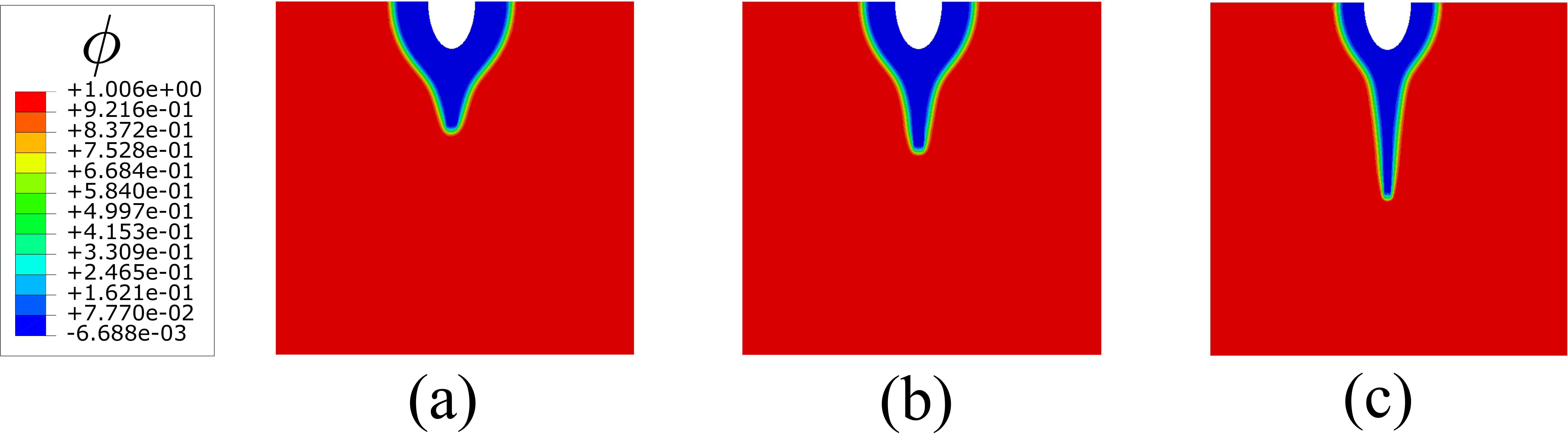}}
\caption{SCC from a semi-elliptical pit. Contours of the phase field after $t=900$ s for a remote displacement of $u^\infty=0.1$ $\mu$m and three selected values of the film stability parameter: (a) $k=2 \times 10^{-4}$, (b) $k=5 \times 10^{-4}$, and (c) $k=1 \times 10^{-3}$.}
\label{fig:Veri_3_5}
\end{figure}

\begin{figure}[H]
\centering
\noindent\makebox[\textwidth]{%
\includegraphics[scale=0.8]{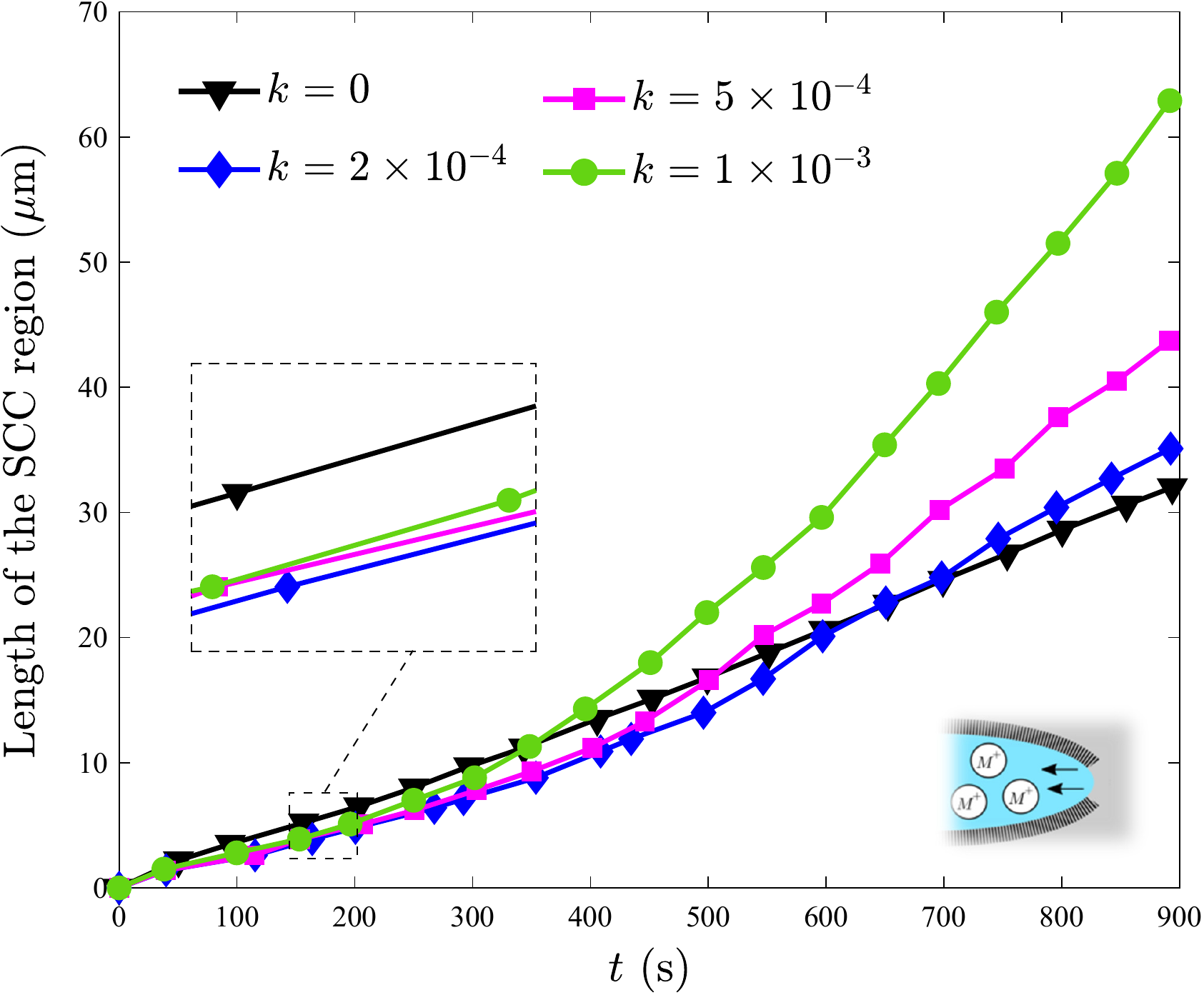}}
\caption{SCC from a semi-elliptical pit. Length of the SCC region as a function of time for a remote displacement of $u^\infty=0.1$ $\mu$m and four selected values of the film stability parameter.}
\label{fig:Veri_3_6}
\end{figure}

The phase field contours obtained after 900 s are shown in Fig. \ref{fig:Veri_3_5}. The impact of $k$ in the SCC morphology is evident; higher $k$ values accelerate the SCC defect growth rate in the pit base relative to corrosion rates at the pit mouth, sharpening the pit and triggering a pit-to-crack transition. Corrosion rates at the tip of the SCC defect are exacerbated as the defect sharpens and plasticity localises; the protective film is weaker in areas of high mechanical straining - see Eq. (\ref{Eq:tf}). The length of the SCC region is shown in Fig. \ref{fig:Veri_3_6} as a function of the film stability parameter $k$, including the case where no film is present ($k=0$). The results reveal that by introducing the repassivation process, the length of the SCC region drops slightly at an early stage, which is in accordance with the fact that repassivation will restrain the SCC process by decreasing the value of interface kinetics coefficient. After a certain time, the size of the SCC region grows faster with increasing $k$ as film rupture occurs and the magnitude of the equivalent plastic strain increases as the defect sharpens, augmenting the interface kinetics coefficient - see Eq. (\ref{Eq:cycle}). 

\begin{figure}[H]
\centering
\noindent\makebox[\textwidth]{%
\includegraphics[scale=0.23]{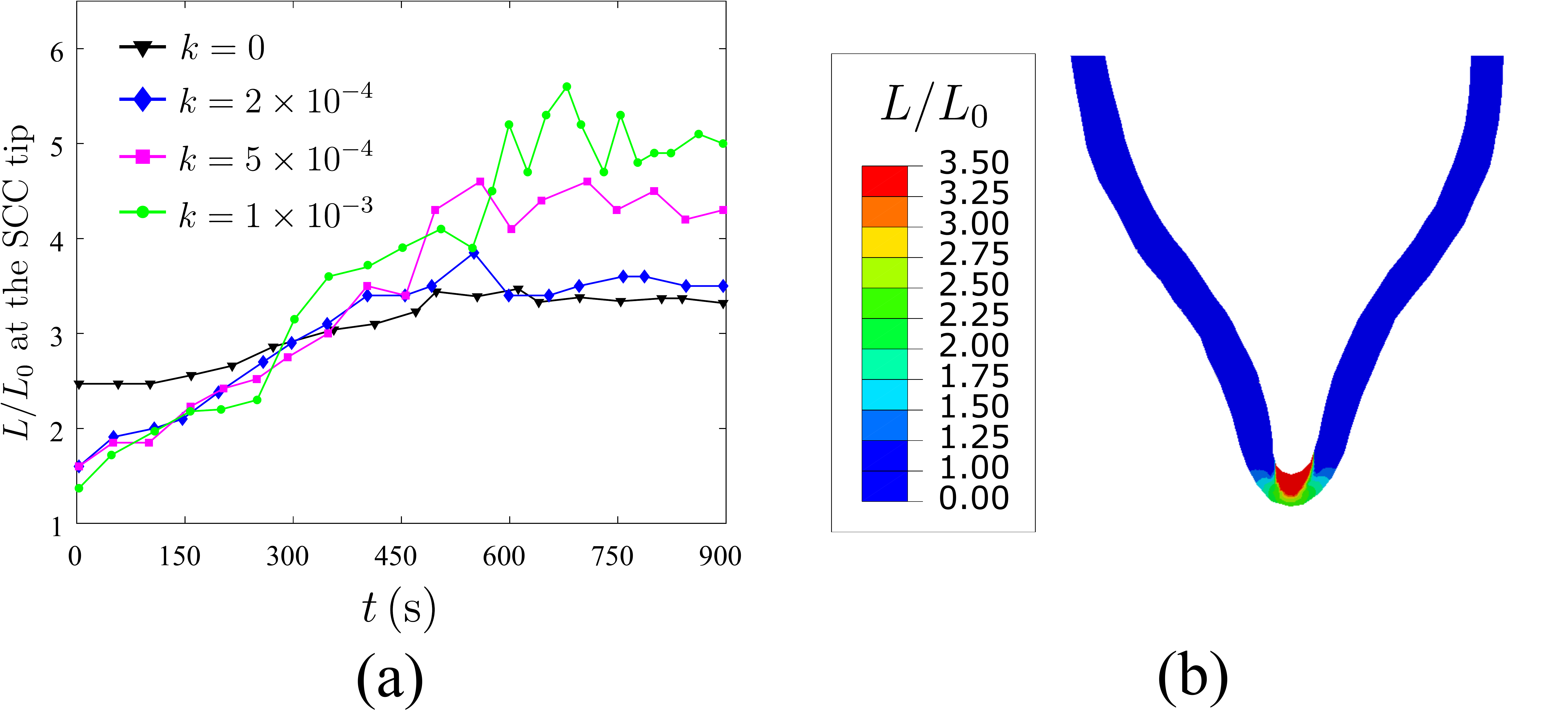}}
\caption{SCC from a semi-elliptical pit, local insight. (a) Evolution of the interface kinetics coefficient at the SCC tip for a remote displacement of $u^\infty=0.1$ $\mu$m and four selected values of the film stability parameter $k$; and (b) spatial contour of $L/L_0$ along the interface for $t=900 \, \mathrm{s}$ and $k=0.0002$.}
\label{fig:Veri_3_7}
\end{figure}

Finally, we gain insight into the model predictions at the local level by quantifying the evolution the interface kinetics coefficient as a function of time and space. We compare the interface kinetics coefficient $L$ at the SCC tip to further clarify the local behaviours. First, Fig. \ref{fig:Veri_3_7}a shows the evolution in time of the normalised interface kinetics coefficient $L/L_0$ at the tip of the SCC defect for different values of the film stability coefficient $k$. The initial value of $L$ is smaller with increasing $k$ due to the effect of film passivation, in agreement with expectations. In all cases, the magnitude of $L/L_0$ increases with time but eventually reaches a plateau. This indicates that mechanical straining dominates the corrosion process. Also, it is observed that SCC damage becomes more localised as the stability of the film $k$ increases. This is showcased in Fig. \ref{fig:Veri_3_7}b, where the spatial contour of $L/L_0$ is shown in the interface region. Higher values of the interface kinetics coefficient are observed at the tip of the pit, where the mechanical fields ($\varepsilon^p$, $\sigma_h$) are larger.

\subsection{SCC driven by the interaction between multiple pits}
\label{sec:casestudy4}

\begin{figure}[H]
\centering
\noindent\makebox[\textwidth]{%
\includegraphics[scale=0.5]{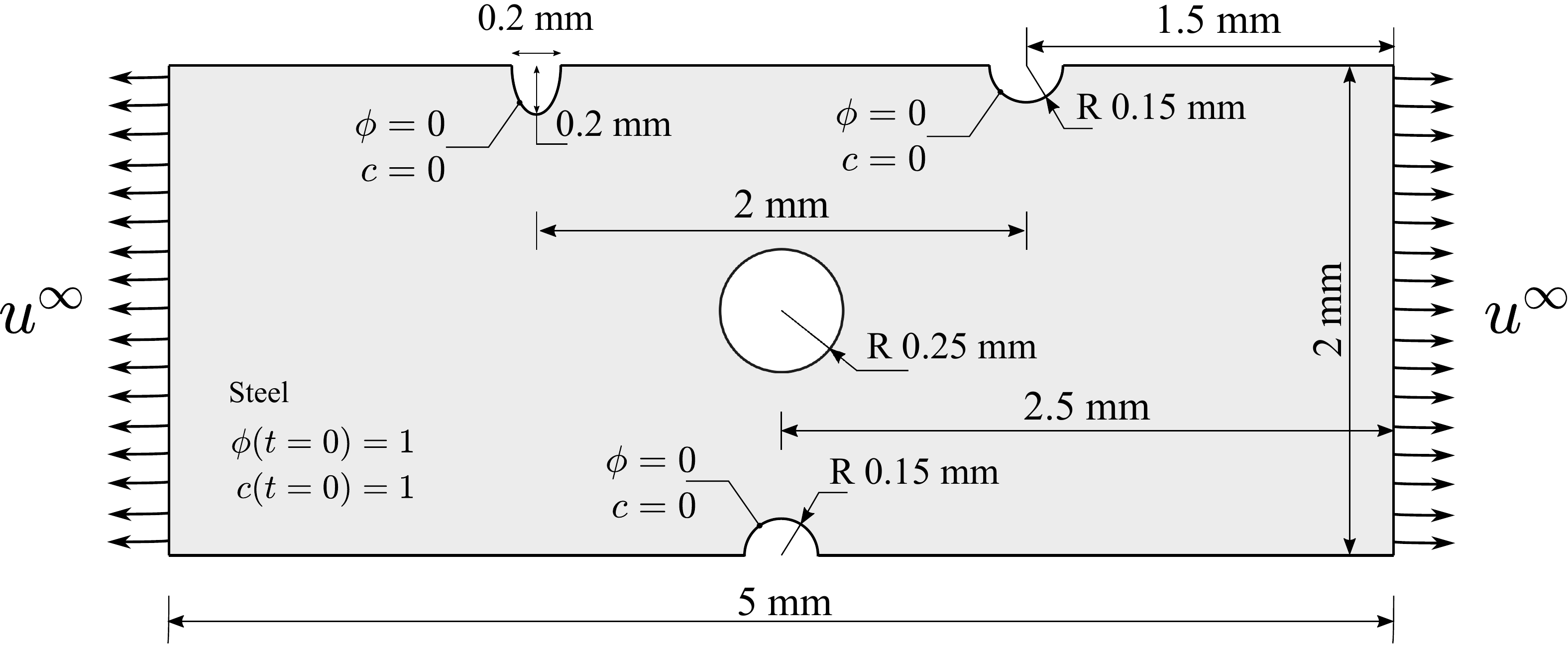}}
\caption{Multiple-pit SCC. Geometry, initial and boundary conditions. The centre hole is not exposed to the environment.}
\label{fig:case_1_1}
\end{figure}

Component failure frequently takes place as a consequence of the interaction between multiple pits. In this section, we demonstrate the capabilities of the model in predicting failure due to the combined action of multiple pits and their coalescence, as well as investigating the role of corrosion parameters in driving the evolution of neighbouring pits. For this purpose, we model a steel plate with an inner defect (not exposed to the environment) and three pre-existing corrosion pits. As shown in Fig. \ref{fig:case_1_1}, one of the existing pits has a semi-ellipsoidal shape with radii 0.1 and 0.2 mm, while the other two pits have a semi-circular profile and the same radius: 0.15 mm. The mechanical load is applied by prescribing a remote horizontal displacement equal to $u^\infty=0.01$ on each side of the plate. The displacement is prescribed at the beginning of the analysis and held constant afterwards (step function).

The corrosion and material parameters considered are those listed in Tables \ref{tab:corro_para} and \ref{tab:mecha_para}, respectively. However, note that a smaller value is assumed for the interface kinetics coefficient, $L_0=1\times 10^{-4} \, \mathrm{mm^2/(N \cdot s)}$, so as to simulate activation-controlled corrosion. Also, the interface thickness is chosen to be equal to $l=0.2 \, \mathrm{mm}$, which is 20 times larger than the characteristic element size. The plate is discretised using 27,450 quadrilateral quadratic plane strain elements with reduced integration. In addition, the presence of a protective film and its rupture is captured by considering a film stability parameter $k=2 \times 10^{-4}$.\\ 

\begin{figure}[H]
\centering
\noindent\makebox[\textwidth]{%
\includegraphics[scale=0.27]{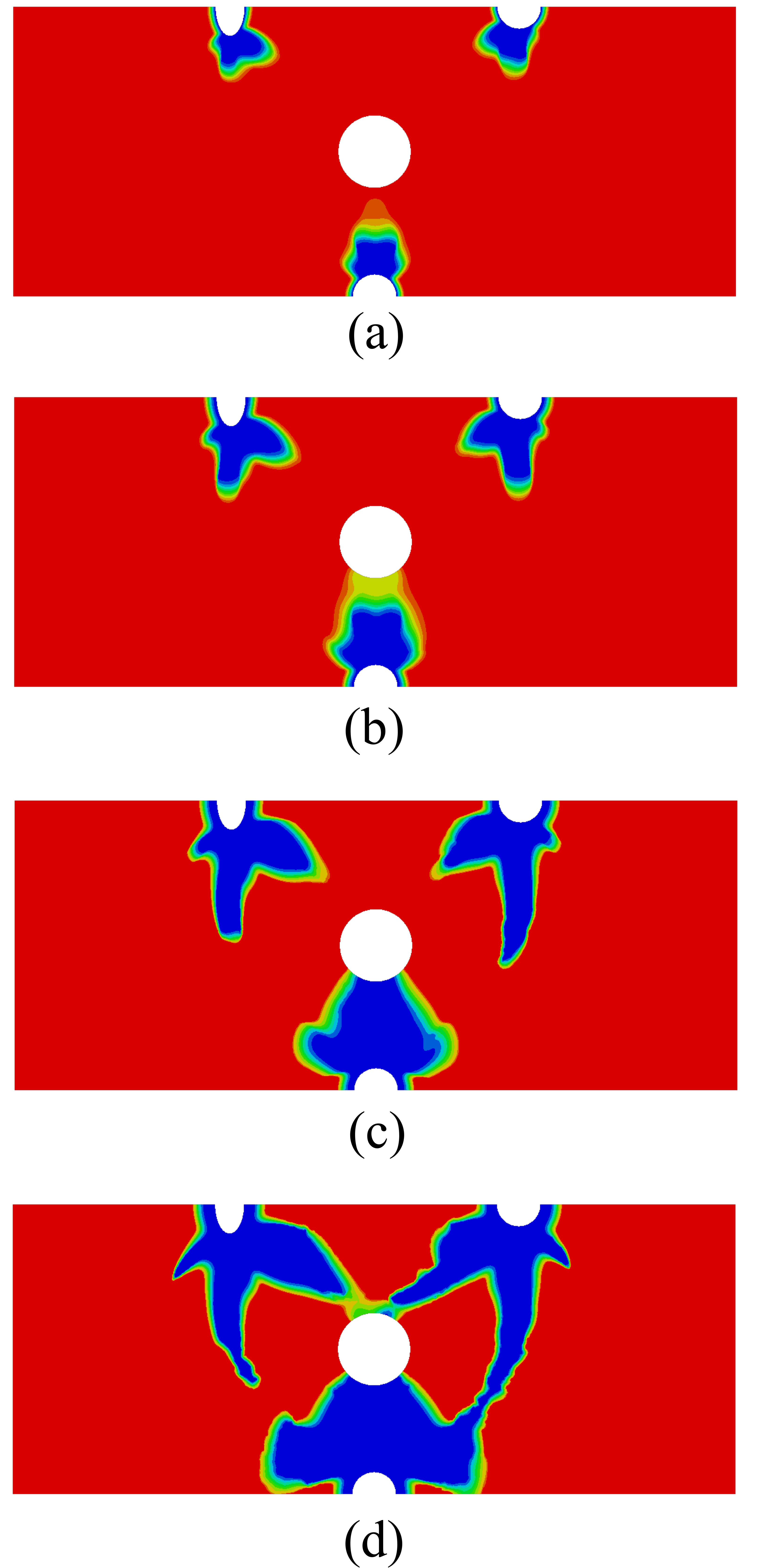}}
\caption{Multiple-pit SCC. Phase field contours at times (a) 0.5 h, (b) 1 h, (c) 2 h, and (d) 4 h. Blue and red colours respectively denote the electrolyte ($\phi=0$) and solid ($\phi=1$) regions.}
\label{fig:case_1_2}
\end{figure}

The SCC process is shown as a function of time in Fig. \ref{fig:case_1_2}, where the blue and red colours correspond to the completely corroded and intact areas, respectively. We can see that the three pits increase in depth (Fig. \ref{fig:case_1_2}a), with the upper two eventually branching (Fig. \ref{fig:case_1_2}b). The bottom pit eventually coalesces with the central circular defect (Fig. \ref{fig:case_1_2}c), while the branches of the upper defects sharpen - one growing toward the central defect and the other one in the direction of the bottom pit. Eventually, after approximately 4 hours, the upper and bottom pits merge and complete failure of the plate is predicted (Fig. \ref{fig:case_1_2}d).

\begin{figure}[H]
\centering
\noindent\makebox[\textwidth]{%
\includegraphics[scale=0.3]{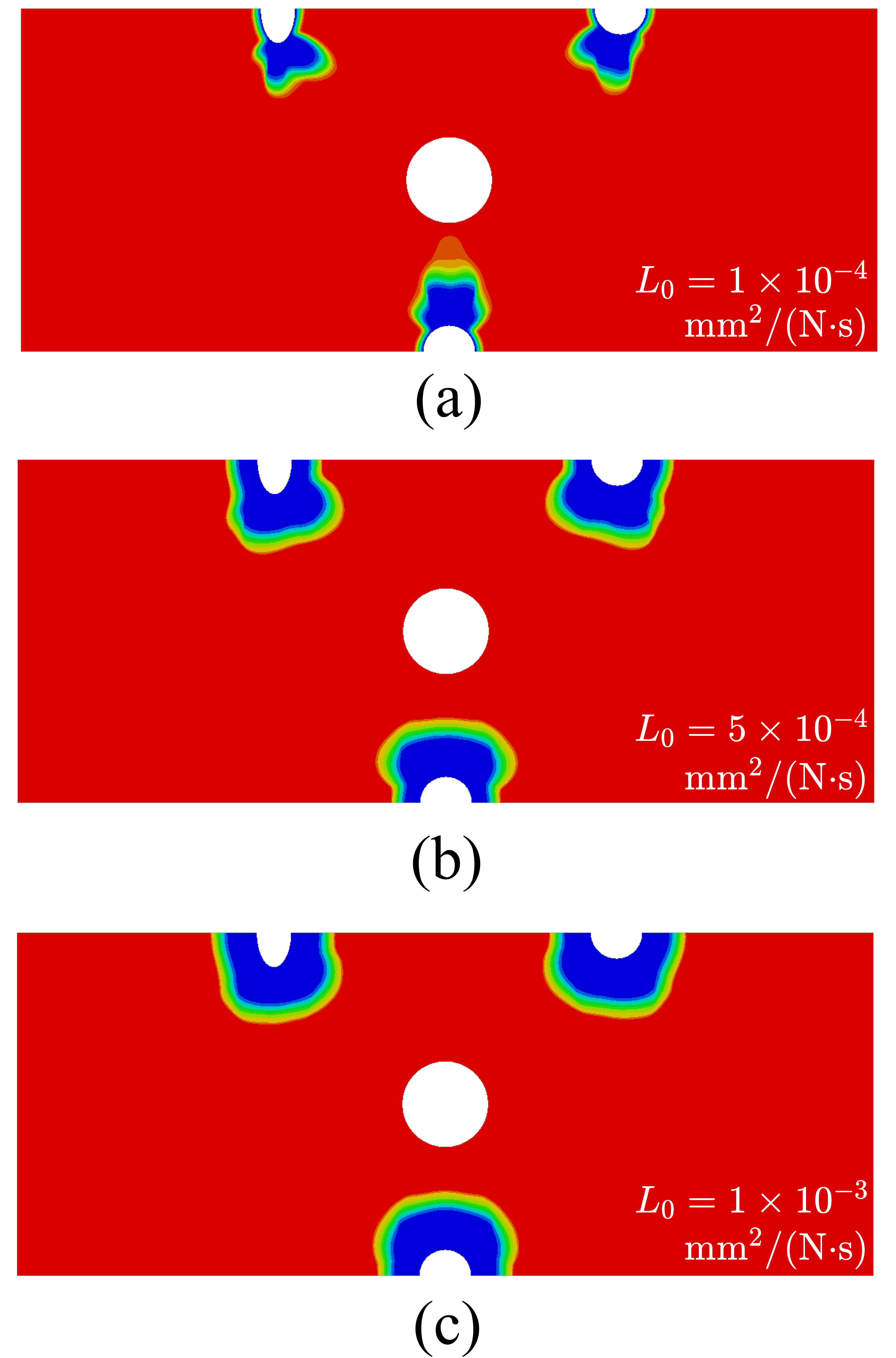}}
\caption{Multiple-pit SCC. Phase field contours after 0.5 h for (a) $L_0=1\times 10^{-4} \, \mathrm{mm^2/(N \cdot s)}$, (b) $L_0=5\times 10^{-4} \, \mathrm{mm^2/(N \cdot s)}$, and (c) $L_0=1\times 10^{-3} \, \mathrm{mm^2/(N \cdot s)}$.}
\label{fig:case_1_3}
\end{figure}

We also compare the SCC morphologies for three different values of $L_0$ to assess the influence of the interface kinetics parameter on the corrosion process. The results are shown in Fig. \ref{fig:case_1_3} for a time of $t$=0.5 h. It can be seen that the pits grow more uniformly when the initial interface kinetics parameter is higher, implying a smaller contribution of the mechanical problem with increasing $L_0$. Mechanical strains and stresses enhance interface kinetics via the mechanochemical term $k_\mathrm{m} \left( \varepsilon^p, \sigma_h \right)$, see (\ref{Eq:Gutman}) and (\ref{Eq:cycle}). However, this also implies that the mechanical contribution facilitates the transition from activation-controlled corrosion to diffusion-controlled corrosion, and when $L$ is sufficiently large such that corrosion is diffusion-limited, corrosion rates become independent of $L$ (see Fig. \ref{fig:Veri_1_3}) and accordingly of $k_\mathrm{m}$. Consequently, for a given value of the remote load, the contribution from the mechanical problem will be larger for lower $L_0$ magnitudes. 

\subsection{Three-dimensional predictions of SCC defect growth}
\label{eq:casestudy5}

\begin{figure}[H]
\centering
\noindent\makebox[\textwidth]{%
\includegraphics[scale=0.7]{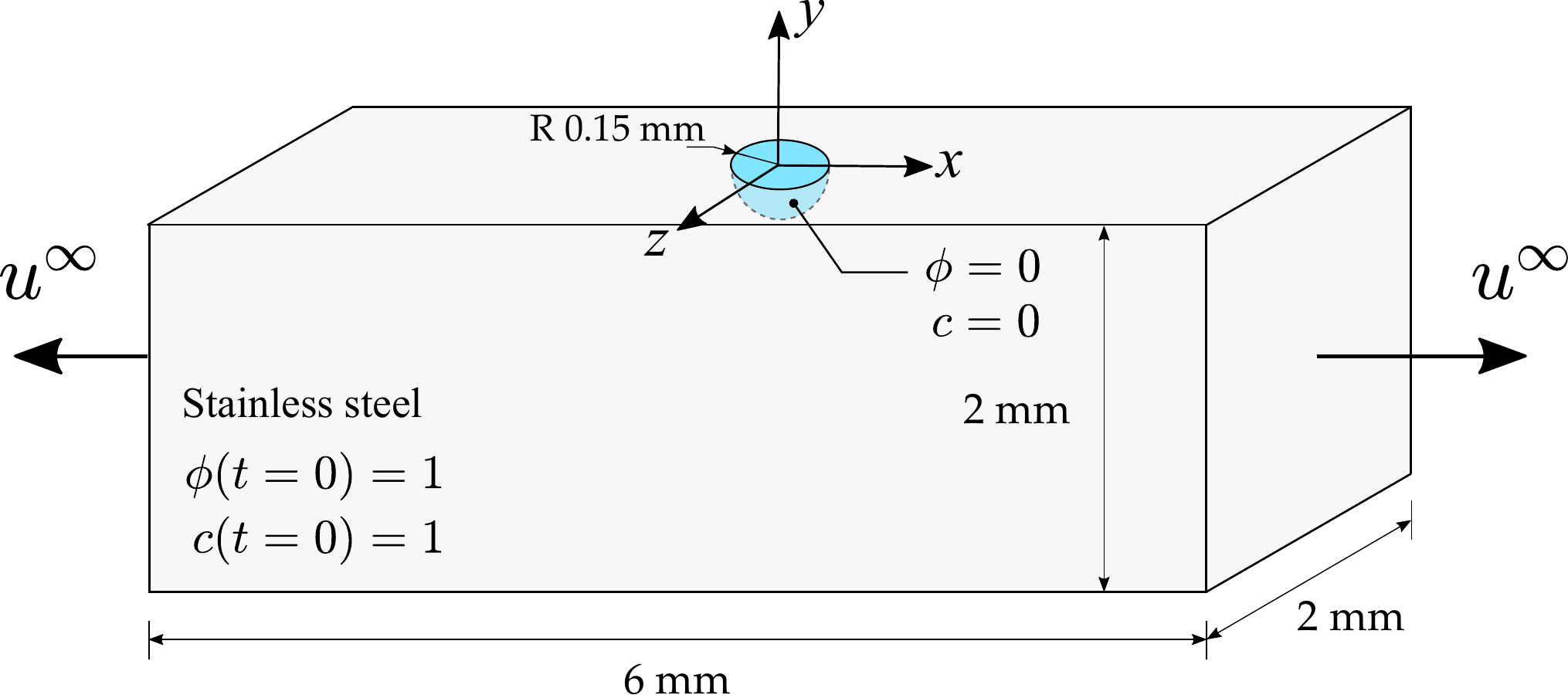}}
\caption{3D pit growth study: geometry and loading configuration. The pit is a hemisphere located at the center of the upper surface of the plate.}
\label{fig:case_2_1}
\end{figure}

Finally, we extend our computational framework to model 3D SCC problems. The goal is to bring new insight into the mechanics of SCC defect growth and demonstrate the capabilities of our model in addressing technologically-relevant case studies. For this purpose, we choose to model a hemispherical pit in a sufficiently large steel plate subjected to tension, aiming to rationalise the observation of cracks initiating at the cross-section of uniaxial tension samples \citep{Turnbull2010}. The pit, of radius 0.15 mm, has its centre located at 3 mm from the loading edge and at 1 mm along the plate thickness. The geometry, initial and boundary conditions are given in Fig. \ref{fig:case_2_1}. The remote displacement, applied along the $x$ axis, is prescribed at the beginning of the numerical experiment and is held constant throughout the analysis; its magnitude equals $u^\infty=0.01$ mm. The corrosion and material parameters adopted are those listed in Tables \ref{tab:corro_para} and \ref{tab:mecha_para} with the exception of an initial kinetics coefficient of $L_0=3\times 10^{-4} \, \mathrm{mm^2/(N \cdot s)}$, to ensure that at $t=0^+$ corrosion is activation-controlled. Also, the presence of a protective film is considered, with $k=2 \times 10^{-4}$. The interface thickness is chosen to be $l=0.4$ mm, which is 20 times larger than the characteristic element size. A total of 356,892 10-node tetrahedral elements are used; the model contains approximately 2.5 million degrees-of-freedom (DOFs). 

\begin{figure}[H]
\centering
\noindent\makebox[\textwidth]{%
\includegraphics[scale=0.18]{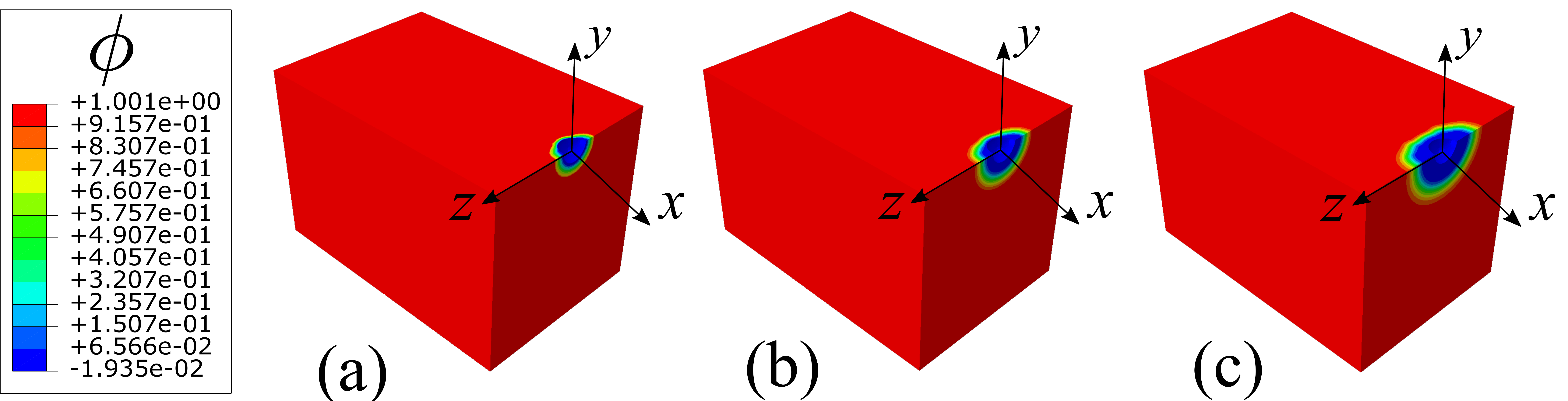}}
\caption{3D pit growth study. Phase field contours after a time of: (a) $t=300 \, \mathrm{s}$, (b) $t=1200 \, \mathrm{s}$, and (c) $t=2100 \, \mathrm{s}$. Blue and red colours respectively denote the electrolyte ($\phi=0$) and solid ($\phi=1$) regions.}
\label{fig:case_2_2}
\end{figure}

\begin{figure}[H]
\centering
\noindent\makebox[\textwidth]{%
\includegraphics[scale=0.25]{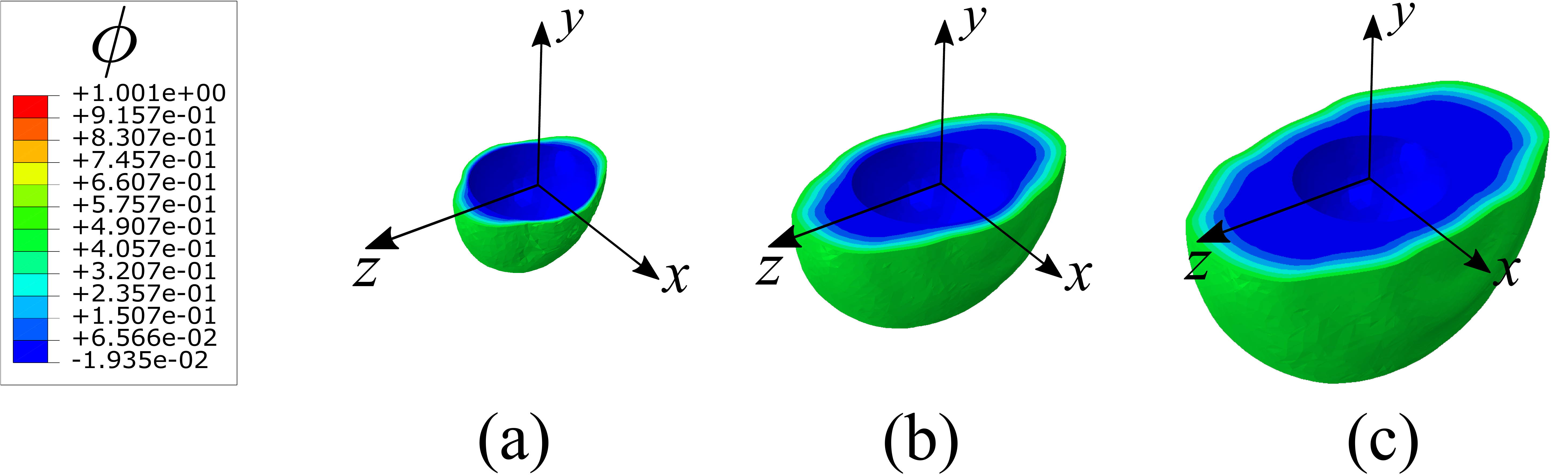}}
\caption{3D pit growth study. Evolution of the SCC defect ($\phi<0.5$) after a time of: (a) $t=300 \, \mathrm{s}$, (b) $t=1200 \, \mathrm{s}$, and (c) $t=2100 \, \mathrm{s}$.}
\label{fig:case_2_3}
\end{figure}

The predicted evolution of the SCC defect is shown in Figs. \ref{fig:case_2_2} and \ref{fig:case_2_3}, in terms of 3D phase field contours. For the sake of a better visualisation, a cut along the $y$-$z$ plane is made in Fig. \ref{fig:case_2_2} while Fig. \ref{fig:case_2_3} shows the SCC region where $\phi < 0.5$. It can be observed that the pit tends to grow more along the $y$ and $z$ directions, relative to the $x$ one. This shape evolution of the pit, going from hemispherical to ellipsoidal, is driven by the influence of the mechanical problem. Plastic straining is larger at the pit base and the regions of the pit mouth closer to the cross-section perpendicular to the applied load. Consequently, the magnitude of $k_\mathrm{m} \left( \varepsilon^p, \sigma_h \right)$ is larger in these regions and the passive film weaker, leading to an increase in the interface kinetics coefficient. This is in agreement with experimental observations of increasing pit depth and the initiation of lateral cracks \citep{Turnbull2001}.  

\begin{figure}[H]
\centering
\noindent\makebox[\textwidth]{%
\includegraphics[scale=0.8]{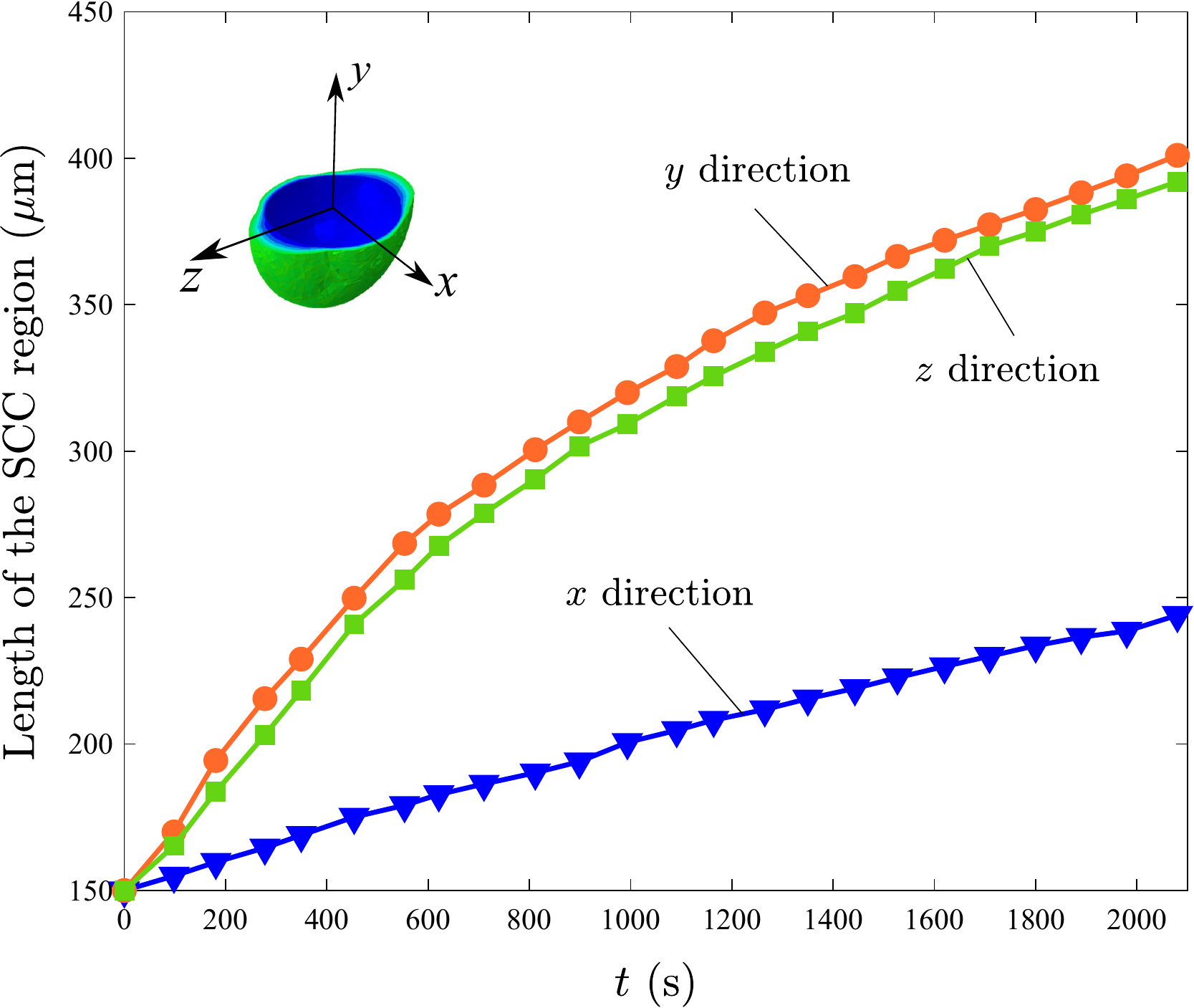}}
\caption{Evolution in time of the pit depth along each of the axes of a Cartesian coordinate axis with its origin at the center of the pit.}
\label{fig:case_2_4}
\end{figure}

The pit depth along the three Cartesian coordinate axes is shown in Fig. \ref{fig:case_2_4} as a function of time. The results show that the pit growth along the $x$ direction is linear, i.e. activation-controlled corrosion. However, the results obtained along the $y$ and $z$ directions transition from linear to parabolic. In other words, the mechanical enhancement of the interface kinetics coefficient $L$ has triggered a transition from activation-controlled to diffusion-controlled corrosion. The role of the mechanical contribution will eventually become negligible as $L$ raises above its saturation value (diffusion-limited corrosion) but those areas where mechanical straining is higher will continue to corrode faster until $L$ reaches its saturation value everywhere, when uniform corrosion will be recovered. 

\section{Concluding remarks}
\label{Sec:Conclusions}

We have presented a new formulation for predicting dissolution-driven pitting and stress corrosion cracking. Its main ingredients are: (i) a KKS-based phase field model for both activation-controlled and diffusion-controlled corrosion, (ii) a mechanics-enhanced interface kinetics law grounded on mechanochemical theory, and (iii) a strain rate governed film rupture and repassivation process according to the FRDR mechanism. A numerical framework is also presented, with displacements, phase field and concentration of dissolved ions as degrees of freedom. The model is benchmarked by addressing five boundary value problems of particular interest. First, we model the paradigmatic pencil electrode test to verify our numerical implementation and examine the model capabilities in predicting the transition from activation-controlled to diffusion-controlled corrosion. Secondly, we reproduce C-ring experiments in a film-free environment to benchmark our mechanics-enhanced corrosion laws against experimental data. Thirdly, SCC from an ellipsoidal pit is predicted under different conditions to quantify the influence of mechanical stresses and their interplay with the film rupture and repassivation process. Fourthly, the interaction between multiple pits is addressed, assessing the competing influences of the coefficients characterising the mechanical, film stability and interface kinetics contributions. Finally, pit growth in a 3D solid is predicted to showcase the capabilities of the model in capturing large scale phenomena and impacting engineering practice. The main findings are:

\begin{itemize}
    \item The interface kinetics coefficient $L$ governs corrosion rates. Corrosion kinetics rise with increasing $L$, going from activation-controlled to diffusion-controlled corrosion, until the process becomes diffusion-limited, when corrosion rates become insensitive to a further augmenting $L$.
    
    \item Mechanical stresses and strains enhance corrosion kinetics, in agreement with experiments, favouring localised damage and triggering a pit-to-crack transition. However, if the mechanical enhancement of $L$ is sufficiently large, the diffusion-controlled limit is reached and the influence of the mechanical contribution reaches a saturation stage. Thus, the influence of mechanical fields is most notable for activation-controlled corrosion.
    
    \item The presence of a passive film exacerbates localisation and the role of mechanics. The effect is more pronounced if the film stability coefficient is large as film rupture will only occur in regions where mechanical straining is high. Also, cracks are more likely to nucleate under activation-controlled corrosion conditions.
    
\end{itemize}

Future developments could involve the extension of the model to consider other cracking mechanisms, including those driven by cathodic reactions. Also, material microstructure and microchemistry can play an important role and should be taken into consideration in many material systems for an accurate prediction of the initial development stages.

\section{Acknowledgements}
\label{Sec:Acknowledgeoffunding}

E. Mart\'{\i}nez-Pa\~neda acknowledges valuable discussions with Z.D. Harris (University of Virginia) and A. Turnbull (National Physical Laboratory). Chuanjie Cui and Rujin Ma acknowledge financial support from National Natural Science Foundation of China under grant No. 51878493. E. Mart\'{\i}nez-Pa\~neda acknowledges financial support from the EPSRC (grants EP/R010161/1 and EP/R017727/1) and from the Royal Commission for the 1851 Exhibition (RF496/2018).



\bibliographystyle{elsarticle-harv}
\bibliography{library}

\begin{thebibliography}{52}
\expandafter\ifx\csname natexlab\endcsname\relax\def\natexlab#1{#1}\fi
\providecommand{\url}[1]{\texttt{#1}}
\providecommand{\href}[2]{#2}
\providecommand{\path}[1]{#1}
\providecommand{\DOIprefix}{doi:}
\providecommand{\ArXivprefix}{arXiv:}
\providecommand{\URLprefix}{URL: }
\providecommand{\Pubmedprefix}{pmid:}
\providecommand{\doi}[1]{\href{http://dx.doi.org/#1}{\path{#1}}}
\providecommand{\Pubmed}[1]{\href{pmid:#1}{\path{#1}}}
\providecommand{\bibinfo}[2]{#2}
\ifx\xfnm\relax \def\xfnm[#1]{\unskip,\space#1}\fi
%Type = Article
\bibitem[{Abubakar et~al.(2015)Abubakar, Akhtar and Arif}]{Abubakar2015}
\bibinfo{author}{Abubakar, A.A.}, \bibinfo{author}{Akhtar, S.S.},
  \bibinfo{author}{Arif, A.F.M.}, \bibinfo{year}{2015}.
\newblock \bibinfo{title}{{Phase field modeling of V2O5 hot corrosion kinetics
  in thermal barrier coatings}}.
\newblock \bibinfo{journal}{Computational Materials Science}
  \bibinfo{volume}{99}, \bibinfo{pages}{105--116}.
%Type = Article
\bibitem[{Anand et~al.(2019)Anand, Mao and Talamini}]{Anand2019}
\bibinfo{author}{Anand, L.}, \bibinfo{author}{Mao, Y.},
  \bibinfo{author}{Talamini, B.}, \bibinfo{year}{2019}.
\newblock \bibinfo{title}{{On modeling fracture of ferritic steels due to
  hydrogen embrittlement}}.
\newblock \bibinfo{journal}{Journal of the Mechanics and Physics of Solids}
  \bibinfo{volume}{122}, \bibinfo{pages}{280--314}.
%Type = Article
\bibitem[{Andresen and Ford(1988)}]{Andresen1988}
\bibinfo{author}{Andresen, P.L.}, \bibinfo{author}{Ford, F.P.},
  \bibinfo{year}{1988}.
\newblock \bibinfo{title}{{Life prediction by mechanistic modeling and system
  monitoring of environmental cracking of iron and nickel alloys in aqueous
  systems}}.
\newblock \bibinfo{journal}{Materials Science and Engineering}
  \bibinfo{volume}{103}, \bibinfo{pages}{167--184}.
%Type = Book
\bibitem[{Biner(2017)}]{Biner2017}
\bibinfo{author}{Biner, S.B.}, \bibinfo{year}{2017}.
\newblock \bibinfo{title}{{Programming Phase-Field Modeling}}.
\newblock \bibinfo{publisher}{Springer}.
%Type = Article
\bibitem[{Bourdin et~al.(2000)Bourdin, Francfort and Marigo}]{Bourdin2000}
\bibinfo{author}{Bourdin, B.}, \bibinfo{author}{Francfort, G.A.},
  \bibinfo{author}{Marigo, J.J.}, \bibinfo{year}{2000}.
\newblock \bibinfo{title}{{Numerical experiments in revisited brittle
  fracture}}.
\newblock \bibinfo{journal}{Journal of the Mechanics and Physics of Solids}
  \bibinfo{volume}{48}, \bibinfo{pages}{797--826}.
%Type = Article
\bibitem[{Carrara et~al.(2020)Carrara, Ambati, Alessi and {De
  Lorenzis}}]{Carrara2020}
\bibinfo{author}{Carrara, P.}, \bibinfo{author}{Ambati, M.},
  \bibinfo{author}{Alessi, R.}, \bibinfo{author}{{De Lorenzis}, L.},
  \bibinfo{year}{2020}.
\newblock \bibinfo{title}{{A framework to model the fatigue behavior of brittle
  materials based on a variational phase-field approach}}.
\newblock \bibinfo{journal}{Computer Methods in Applied Mechanics and
  Engineering} \bibinfo{volume}{361}, \bibinfo{pages}{112731}.
%Type = Article
\bibitem[{Chen and Bobaru(2015)}]{Chen2015}
\bibinfo{author}{Chen, Z.}, \bibinfo{author}{Bobaru, F.}, \bibinfo{year}{2015}.
\newblock \bibinfo{title}{{Peridynamic modeling of pitting corrosion damage}}.
\newblock \bibinfo{journal}{Journal of the Mechanics and Physics of Solids}
  \bibinfo{volume}{78}, \bibinfo{pages}{352--381}.
%Type = Article
\bibitem[{Crane and Gangloff(2016)}]{Crane2016}
\bibinfo{author}{Crane, C.B.}, \bibinfo{author}{Gangloff, R.P.},
  \bibinfo{year}{2016}.
\newblock \bibinfo{title}{{Stress corrosion cracking of Al-Mg alloy 5083
  sensitized at low temperature}}.
\newblock \bibinfo{journal}{Corrosion} \bibinfo{volume}{72},
  \bibinfo{pages}{221--241}.
%Type = Article
\bibitem[{Dai et~al.(2020)Dai, Shi, Guo and Chen}]{Dai2020}
\bibinfo{author}{Dai, H.}, \bibinfo{author}{Shi, S.}, \bibinfo{author}{Guo,
  C.}, \bibinfo{author}{Chen, X.}, \bibinfo{year}{2020}.
\newblock \bibinfo{title}{{Pits formation and stress corrosion cracking
  behavior of Q345R in hydrofluoric acid}}.
\newblock \bibinfo{journal}{Corrosion Science} \bibinfo{volume}{166},
  \bibinfo{pages}{108443}.
%Type = Article
\bibitem[{Duda et~al.(2018)Duda, Ciarbonetti, Toro and Huespe}]{Duda2018}
\bibinfo{author}{Duda, F.P.}, \bibinfo{author}{Ciarbonetti, A.},
  \bibinfo{author}{Toro, S.}, \bibinfo{author}{Huespe, A.E.},
  \bibinfo{year}{2018}.
\newblock \bibinfo{title}{{A phase-field model for solute-assisted brittle
  fracture in elastic-plastic solids}}.
\newblock \bibinfo{journal}{International Journal of Plasticity}
  \bibinfo{volume}{102}, \bibinfo{pages}{16--40}.
%Type = Article
\bibitem[{Duddu et~al.(2016)Duddu, Kota and Qidwai}]{Duddu2016}
\bibinfo{author}{Duddu, R.}, \bibinfo{author}{Kota, N.},
  \bibinfo{author}{Qidwai, S.M.}, \bibinfo{year}{2016}.
\newblock \bibinfo{title}{{An Extended Finite Element Method Based Approach for
  Modeling Crevice and Pitting Corrosion}}.
\newblock \bibinfo{journal}{Journal of Applied Mechanics} \bibinfo{volume}{83},
  \bibinfo{pages}{1--10}.
%Type = Article
\bibitem[{Ernst and Newman(2002)}]{Ernst2002}
\bibinfo{author}{Ernst, P.}, \bibinfo{author}{Newman, R.C.},
  \bibinfo{year}{2002}.
\newblock \bibinfo{title}{{Pit growth studies in stainless steel foils. I.
  Introduction and pit growth kinetics}}.
\newblock \bibinfo{journal}{Corrosion Science} \bibinfo{volume}{44},
  \bibinfo{pages}{927--941}.
%Type = Article
\bibitem[{Gao et~al.(2020)Gao, Ju, Duddu and Li}]{Gao2020}
\bibinfo{author}{Gao, H.}, \bibinfo{author}{Ju, L.}, \bibinfo{author}{Duddu,
  R.}, \bibinfo{author}{Li, H.}, \bibinfo{year}{2020}.
\newblock \bibinfo{title}{{An efficient second-order linear scheme for the
  phase field model of corrosive dissolution}}.
\newblock \bibinfo{journal}{Journal of Computational and Applied Mathematics}
  \bibinfo{volume}{367}, \bibinfo{pages}{112472}.
%Type = Book
\bibitem[{Gurtin et~al.(2010)Gurtin, Fried and Anand}]{Gurtin2010}
\bibinfo{author}{Gurtin, M.E.}, \bibinfo{author}{Fried, E.},
  \bibinfo{author}{Anand, L.}, \bibinfo{year}{2010}.
\newblock \bibinfo{title}{{The Mechanics and Thermodynamics of continua}}.
\newblock \bibinfo{publisher}{Cambridge University Press},
  \bibinfo{address}{Cambridge, UK}.
%Type = Book
\bibitem[{Gutman(1998)}]{Gutman1998}
\bibinfo{author}{Gutman, E.M.}, \bibinfo{year}{1998}.
\newblock \bibinfo{title}{{Mechanochemistry of materials}}.
\newblock \bibinfo{publisher}{Cambridge International Science Publishing},
  \bibinfo{address}{Cambridge, UK}.
%Type = Article
\bibitem[{Gutman(2007)}]{Gutman2007}
\bibinfo{author}{Gutman, E.M.}, \bibinfo{year}{2007}.
\newblock \bibinfo{title}{{An inconsistency in "film rupture model" of stress
  corrosion cracking}}.
\newblock \bibinfo{journal}{Corrosion Science} \bibinfo{volume}{49},
  \bibinfo{pages}{2289--2302}.
%Type = Article
\bibitem[{Hirshikesh et~al.(2019)Hirshikesh, Natarajan, Annabattula and
  Mart{\'{i}}nez-Pa{\~{n}}eda}]{CPB2019}
\bibinfo{author}{Hirshikesh}, \bibinfo{author}{Natarajan, S.},
  \bibinfo{author}{Annabattula, R.K.},
  \bibinfo{author}{Mart{\'{i}}nez-Pa{\~{n}}eda, E.}, \bibinfo{year}{2019}.
\newblock \bibinfo{title}{{Phase field modelling of crack propagation in
  functionally graded materials}}.
\newblock \bibinfo{journal}{Composites Part B: Engineering}
  \bibinfo{volume}{169}, \bibinfo{pages}{239--248}.
%Type = Article
\bibitem[{Jivkov(2004)}]{Jivkov2004}
\bibinfo{author}{Jivkov, A.P.}, \bibinfo{year}{2004}.
\newblock \bibinfo{title}{{Strain-induced passivity breakdown in corrosion
  crack initiation}}.
\newblock \bibinfo{journal}{Theoretical and Applied Fracture Mechanics}
  \bibinfo{volume}{42}, \bibinfo{pages}{43--52}.
%Type = Article
\bibitem[{Kim et~al.(1999)Kim, Kim and Suzuki}]{Kim1999}
\bibinfo{author}{Kim, S.G.}, \bibinfo{author}{Kim, W.T.},
  \bibinfo{author}{Suzuki, T.}, \bibinfo{year}{1999}.
\newblock \bibinfo{title}{{Phase-field model for binary alloys}}.
\newblock \bibinfo{journal}{Physical Review E - Statistical Physics, Plasmas,
  Fluids, and Related Interdisciplinary Topics} \bibinfo{volume}{60},
  \bibinfo{pages}{7186--7197}.
%Type = Article
\bibitem[{Kristensen and Mart{\'{i}}nez-Pa{\~{n}}eda(2020)}]{TAFM2020}
\bibinfo{author}{Kristensen, P.K.},
  \bibinfo{author}{Mart{\'{i}}nez-Pa{\~{n}}eda, E.}, \bibinfo{year}{2020}.
\newblock \bibinfo{title}{{Phase field fracture modelling using quasi-Newton
  methods and a new adaptive step scheme}}.
\newblock \bibinfo{journal}{Theoretical and Applied Fracture Mechanics}
  \bibinfo{volume}{107}, \bibinfo{pages}{102446}.
%Type = Article
\bibitem[{Kristensen et~al.(2020)Kristensen, Niordson and
  Mart{\'{i}}nez-Pa{\~{n}}eda}]{JMPS2020}
\bibinfo{author}{Kristensen, P.K.}, \bibinfo{author}{Niordson, C.F.},
  \bibinfo{author}{Mart{\'{i}}nez-Pa{\~{n}}eda, E.}, \bibinfo{year}{2020}.
\newblock \bibinfo{title}{{A phase field model for elastic-gradient-plastic
  solids undergoing hydrogen embrittlement}}.
\newblock \bibinfo{journal}{Journal of the Mechanics and Physics of Solids}
  \bibinfo{volume}{143}, \bibinfo{pages}{104093}.
%Type = Article
\bibitem[{Larrosa et~al.(2018)Larrosa, Akid and Ainsworth}]{Larrosa2018a}
\bibinfo{author}{Larrosa, N.O.}, \bibinfo{author}{Akid, R.},
  \bibinfo{author}{Ainsworth, R.A.}, \bibinfo{year}{2018}.
\newblock \bibinfo{title}{{Corrosion-fatigue: a review of damage tolerance
  models}}.
\newblock \bibinfo{journal}{International Materials Reviews}
  \bibinfo{volume}{63}, \bibinfo{pages}{283--308}.
%Type = Article
\bibitem[{Liang et~al.(2019)Liang, Guo, Liu, Yang and Li}]{Liang2019}
\bibinfo{author}{Liang, G.}, \bibinfo{author}{Guo, H.}, \bibinfo{author}{Liu,
  Y.}, \bibinfo{author}{Yang, D.}, \bibinfo{author}{Li, S.},
  \bibinfo{year}{2019}.
\newblock \bibinfo{title}{{A comparative study on tensile behavior of welded
  T-stub joints using Q345 normal steel and Q690 high strength steel under bolt
  preloading cases}}.
\newblock \bibinfo{journal}{Thin-Walled Structures} \bibinfo{volume}{137},
  \bibinfo{pages}{271--283}.
%Type = Article
\bibitem[{Lo et~al.(2019)Lo, Borden, Ravi-Chandar and Landis}]{Lo2019}
\bibinfo{author}{Lo, Y.S.}, \bibinfo{author}{Borden, M.J.},
  \bibinfo{author}{Ravi-Chandar, K.}, \bibinfo{author}{Landis, C.M.},
  \bibinfo{year}{2019}.
\newblock \bibinfo{title}{{A phase-field model for fatigue crack growth}}.
\newblock \bibinfo{journal}{Journal of the Mechanics and Physics of Solids}
  \bibinfo{volume}{132}, \bibinfo{pages}{103684}.
%Type = Article
\bibitem[{Lynch(1988)}]{Lynch1988}
\bibinfo{author}{Lynch, S.P.}, \bibinfo{year}{1988}.
\newblock \bibinfo{title}{{Environmentally assisted cracking: Overview of
  evidence for an adsorption-induced localised-slip process}}.
\newblock \bibinfo{journal}{Acta Metallurgica} \bibinfo{volume}{36},
  \bibinfo{pages}{2639--2661}.
%Type = Article
\bibitem[{Macdonald(1999)}]{Macdonald1999}
\bibinfo{author}{Macdonald, D.D.}, \bibinfo{year}{1999}.
\newblock \bibinfo{title}{{Passivity - the key to our metals-based
  civilization}}.
\newblock \bibinfo{journal}{Pure and Applied Chemistry} \bibinfo{volume}{71},
  \bibinfo{pages}{951--978}.
%Type = Article
\bibitem[{Mai and Soghrati(2017)}]{Mai2017}
\bibinfo{author}{Mai, W.}, \bibinfo{author}{Soghrati, S.},
  \bibinfo{year}{2017}.
\newblock \bibinfo{title}{{A phase field model for simulating the stress
  corrosion cracking initiated from pits}}.
\newblock \bibinfo{journal}{Corrosion Science} \bibinfo{volume}{125},
  \bibinfo{pages}{87--98}.
%Type = Article
\bibitem[{Mai et~al.(2016)Mai, Soghrati and Buchheit}]{Mai2016}
\bibinfo{author}{Mai, W.}, \bibinfo{author}{Soghrati, S.},
  \bibinfo{author}{Buchheit, R.G.}, \bibinfo{year}{2016}.
\newblock \bibinfo{title}{{A phase field model for simulating the pitting
  corrosion}}.
\newblock \bibinfo{journal}{Corrosion Science} \bibinfo{volume}{110},
  \bibinfo{pages}{157--166}.
%Type = Article
\bibitem[{Mart{\'{i}}nez-Pa{\~{n}}eda et~al.(2019)Mart{\'{i}}nez-Pa{\~{n}}eda,
  Deshpande, Niordson and Fleck}]{JMPS2019}
\bibinfo{author}{Mart{\'{i}}nez-Pa{\~{n}}eda, E.}, \bibinfo{author}{Deshpande,
  V.S.}, \bibinfo{author}{Niordson, C.F.}, \bibinfo{author}{Fleck, N.A.},
  \bibinfo{year}{2019}.
\newblock \bibinfo{title}{{The role of plastic strain gradients in the crack
  growth resistance of metals}}.
\newblock \bibinfo{journal}{Journal of the Mechanics and Physics of Solids}
  \bibinfo{volume}{126}, \bibinfo{pages}{136--150}.
%Type = Article
\bibitem[{Mart{\'{i}}nez-Pa{\~{n}}eda and Fleck(2019)}]{EJMAS2019}
\bibinfo{author}{Mart{\'{i}}nez-Pa{\~{n}}eda, E.}, \bibinfo{author}{Fleck,
  N.A.}, \bibinfo{year}{2019}.
\newblock \bibinfo{title}{{Mode I crack tip fields: Strain gradient plasticity
  theory versus J2 flow theory}}.
\newblock \bibinfo{journal}{European Journal of Mechanics - A/Solids}
  \bibinfo{volume}{75}, \bibinfo{pages}{381--388}.
%Type = Article
\bibitem[{Mart{\'{i}}nez-Pa{\~{n}}eda et~al.(2018)Mart{\'{i}}nez-Pa{\~{n}}eda,
  Golahmar and Niordson}]{CMAME2018}
\bibinfo{author}{Mart{\'{i}}nez-Pa{\~{n}}eda, E.}, \bibinfo{author}{Golahmar,
  A.}, \bibinfo{author}{Niordson, C.F.}, \bibinfo{year}{2018}.
\newblock \bibinfo{title}{{A phase field formulation for hydrogen assisted
  cracking}}.
\newblock \bibinfo{journal}{Computer Methods in Applied Mechanics and
  Engineering} \bibinfo{volume}{342}, \bibinfo{pages}{742--761}.
%Type = Article
\bibitem[{Nguyen et~al.(2017a)Nguyen, Bolivar, R{\'{e}}thor{\'{e}}, Baietto and
  Fregonese}]{Nguyen2017a}
\bibinfo{author}{Nguyen, T.T.}, \bibinfo{author}{Bolivar, J.},
  \bibinfo{author}{R{\'{e}}thor{\'{e}}, J.}, \bibinfo{author}{Baietto, M.C.},
  \bibinfo{author}{Fregonese, M.}, \bibinfo{year}{2017}a.
\newblock \bibinfo{title}{{A phase field method for modeling stress corrosion
  crack propagation in a nickel base alloy}}.
\newblock \bibinfo{journal}{International Journal of Solids and Structures}
  \bibinfo{volume}{112}, \bibinfo{pages}{65--82}.
%Type = Article
\bibitem[{Nguyen et~al.(2018)Nguyen, Bolivar, Shi, R{\'{e}}thor{\'{e}}, King,
  Fregonese, Adrien, Buffiere and Baietto}]{Nguyen2018}
\bibinfo{author}{Nguyen, T.T.}, \bibinfo{author}{Bolivar, J.},
  \bibinfo{author}{Shi, Y.}, \bibinfo{author}{R{\'{e}}thor{\'{e}}, J.},
  \bibinfo{author}{King, A.}, \bibinfo{author}{Fregonese, M.},
  \bibinfo{author}{Adrien, J.}, \bibinfo{author}{Buffiere, J.Y.},
  \bibinfo{author}{Baietto, M.C.}, \bibinfo{year}{2018}.
\newblock \bibinfo{title}{{A phase field method for modeling anodic dissolution
  induced stress corrosion crack propagation}}.
\newblock \bibinfo{journal}{Corrosion Science} \bibinfo{volume}{132},
  \bibinfo{pages}{146--160}.
%Type = Article
\bibitem[{Nguyen et~al.(2017b)Nguyen, R{\'{e}}thor{\'{e}}, Baietto, Bolivar,
  Fregonese and Bordas}]{Nguyen2017b}
\bibinfo{author}{Nguyen, T.T.}, \bibinfo{author}{R{\'{e}}thor{\'{e}}, J.},
  \bibinfo{author}{Baietto, M.C.}, \bibinfo{author}{Bolivar, J.},
  \bibinfo{author}{Fregonese, M.}, \bibinfo{author}{Bordas, S.P.},
  \bibinfo{year}{2017}b.
\newblock \bibinfo{title}{{Modeling of inter- and transgranular stress
  corrosion crack propagation in polycrystalline material by using phase field
  method}}.
\newblock \bibinfo{journal}{Journal of the Mechanical Behavior of Materials}
  \bibinfo{volume}{26}, \bibinfo{pages}{181--191}.
%Type = Article
\bibitem[{Oriani(1972)}]{Oriani1972}
\bibinfo{author}{Oriani, R.A.}, \bibinfo{year}{1972}.
\newblock \bibinfo{title}{{A mechanistic theory of hydrogen embrittlement of
  steels}}.
\newblock \bibinfo{journal}{Berichte der Bunsengesellschaft f{\"{u}}r
  physikalische Chemie} \bibinfo{volume}{76}, \bibinfo{pages}{848--857}.
%Type = Article
\bibitem[{Papazafeiropoulos et~al.(2017)Papazafeiropoulos, Mu{\~{n}}iz-Calvente
  and Mart{\'{i}}nez-Pa{\~{n}}eda}]{AES2017}
\bibinfo{author}{Papazafeiropoulos, G.}, \bibinfo{author}{Mu{\~{n}}iz-Calvente,
  M.}, \bibinfo{author}{Mart{\'{i}}nez-Pa{\~{n}}eda, E.}, \bibinfo{year}{2017}.
\newblock \bibinfo{title}{{Abaqus2Matlab: A suitable tool for finite element
  post-processing}}.
\newblock \bibinfo{journal}{Advances in Engineering Software}
  \bibinfo{volume}{105}, \bibinfo{pages}{9--16}.
%Type = Article
\bibitem[{Parkins(1987)}]{Parkins1987}
\bibinfo{author}{Parkins, R.N.}, \bibinfo{year}{1987}.
\newblock \bibinfo{title}{{Factors Influencing Stress Corrosion Crack Growth
  Kinetics.}}
\newblock \bibinfo{journal}{Corrosion} \bibinfo{volume}{43},
  \bibinfo{pages}{130--139}.
%Type = Article
\bibitem[{Parkins(1996)}]{Parkins1996}
\bibinfo{author}{Parkins, R.N.}, \bibinfo{year}{1996}.
\newblock \bibinfo{title}{{Mechanistic Aspects of Intergranular Stress
  Corrosion Cracking of Ferritic Steels}}.
\newblock \bibinfo{journal}{Corrosion} \bibinfo{volume}{52},
  \bibinfo{pages}{363--374}.
%Type = Book
\bibitem[{Raja and Shoji(2011)}]{Raja2011}
\bibinfo{author}{Raja, V.S.}, \bibinfo{author}{Shoji, T.},
  \bibinfo{year}{2011}.
\newblock \bibinfo{title}{{Stress corrosion cracking: theory and practice}}.
\newblock \bibinfo{publisher}{Woodhead Publishing Limited},
  \bibinfo{address}{Cambridge}.
%Type = Article
\bibitem[{Scheiner and Hellmich(2007)}]{Scheiner2007}
\bibinfo{author}{Scheiner, S.}, \bibinfo{author}{Hellmich, C.},
  \bibinfo{year}{2007}.
\newblock \bibinfo{title}{{Stable pitting corrosion of stainless steel as
  diffusion-controlled dissolution process with a sharp moving electrode
  boundary}}.
\newblock \bibinfo{journal}{Corrosion Science} \bibinfo{volume}{49},
  \bibinfo{pages}{319--346}.
%Type = Article
\bibitem[{Scully(1975)}]{Scully1975}
\bibinfo{author}{Scully, J.C.}, \bibinfo{year}{1975}.
\newblock \bibinfo{title}{{Stress corrosion crack propagation: A constant
  charge criterion}}.
\newblock \bibinfo{journal}{Corrosion Science} \bibinfo{volume}{15},
  \bibinfo{pages}{207--224}.
%Type = Article
\bibitem[{Scully(1980)}]{Scully1980}
\bibinfo{author}{Scully, J.C.}, \bibinfo{year}{1980}.
\newblock \bibinfo{title}{{The interaction of strain-rate and repassivation
  rate in stress corrosion crack propagation}}.
\newblock \bibinfo{journal}{Corrosion Science} \bibinfo{volume}{20},
  \bibinfo{pages}{997--1016}.
%Type = Article
\bibitem[{Serebrinsky et~al.(2004)Serebrinsky, Carter and
  Ortiz}]{Serebrinsky2004}
\bibinfo{author}{Serebrinsky, S.}, \bibinfo{author}{Carter, E.A.},
  \bibinfo{author}{Ortiz, M.}, \bibinfo{year}{2004}.
\newblock \bibinfo{title}{{A quantum-mechanically informed continuum model of
  hydrogen embrittlement}}.
\newblock \bibinfo{journal}{Journal of the Mechanics and Physics of Solids}
  \bibinfo{volume}{52}, \bibinfo{pages}{2403--2430}.
%Type = Article
\bibitem[{Sofronis and Birnbaum(1995)}]{Sofronis1995}
\bibinfo{author}{Sofronis, P.}, \bibinfo{author}{Birnbaum, H.},
  \bibinfo{year}{1995}.
\newblock \bibinfo{title}{{Mechanics of the hydrogen-dislocation-impurity
  interactions—I. Increasing shear modulus}}.
\newblock \bibinfo{journal}{Journal of the Mechanics and Physics of Solids}
  \bibinfo{volume}{43}, \bibinfo{pages}{49--90}.
%Type = Article
\bibitem[{St{\aa}hle and Hansen(2015)}]{Stahle2015}
\bibinfo{author}{St{\aa}hle, P.}, \bibinfo{author}{Hansen, E.},
  \bibinfo{year}{2015}.
\newblock \bibinfo{title}{{Phase field modelling of stress corrosion}}.
\newblock \bibinfo{journal}{Engineering Failure Analysis} \bibinfo{volume}{47},
  \bibinfo{pages}{241--251}.
%Type = Article
\bibitem[{Sun et~al.(2014)Sun, Wang, Wu and Liu}]{Sun2014}
\bibinfo{author}{Sun, W.}, \bibinfo{author}{Wang, L.}, \bibinfo{author}{Wu,
  T.}, \bibinfo{author}{Liu, G.}, \bibinfo{year}{2014}.
\newblock \bibinfo{title}{{An arbitrary Lagrangian-Eulerian model for modelling
  the time-dependent evolution of crevice corrosion}}.
\newblock \bibinfo{journal}{Corrosion Science} \bibinfo{volume}{78},
  \bibinfo{pages}{233--243}.
%Type = Article
\bibitem[{Tann{\'{e}} et~al.(2018)Tann{\'{e}}, Li, Bourdin, Marigo and
  Maurini}]{Tanne2018}
\bibinfo{author}{Tann{\'{e}}, E.}, \bibinfo{author}{Li, T.},
  \bibinfo{author}{Bourdin, B.}, \bibinfo{author}{Marigo, J.J.},
  \bibinfo{author}{Maurini, C.}, \bibinfo{year}{2018}.
\newblock \bibinfo{title}{{Crack nucleation in variational phase-field models
  of brittle fracture}}.
\newblock \bibinfo{journal}{Journal of the Mechanics and Physics of Solids}
  \bibinfo{volume}{110}, \bibinfo{pages}{80--99}.
%Type = Phdthesis
\bibitem[{Toshniwal(2019)}]{Toshniwal2019}
\bibinfo{author}{Toshniwal, D.}, \bibinfo{year}{2019}.
\newblock \bibinfo{title}{{Isogeometric Analysis : Study of non-uniform degree
  and unstructured splines , and application to phase field modeling of
  corrosion}}.
\newblock Ph.D. thesis. The University of Texas at Austin.
%Type = Article
\bibitem[{Turnbull(1993)}]{Turnbull1993}
\bibinfo{author}{Turnbull, A.}, \bibinfo{year}{1993}.
\newblock \bibinfo{title}{{Modelling of environment assisted cracking}}.
\newblock \bibinfo{journal}{Corrosion Science} \bibinfo{volume}{34},
  \bibinfo{pages}{921--960}.
%Type = Article
\bibitem[{Turnbull(2001)}]{Turnbull2001}
\bibinfo{author}{Turnbull, A.}, \bibinfo{year}{2001}.
\newblock \bibinfo{title}{{Modeling of the chemistry and electrochemistry in
  cracks - A review}}.
\newblock \bibinfo{journal}{Corrosion} \bibinfo{volume}{57},
  \bibinfo{pages}{175--188}.
%Type = Article
\bibitem[{Turnbull et~al.(2010)Turnbull, Wright and Crocker}]{Turnbull2010}
\bibinfo{author}{Turnbull, A.}, \bibinfo{author}{Wright, L.},
  \bibinfo{author}{Crocker, L.}, \bibinfo{year}{2010}.
\newblock \bibinfo{title}{{New insight into the pit-to-crack transition from
  finite element analysis of the stress and strain distribution around a
  corrosion pit}}.
\newblock \bibinfo{journal}{Corrosion Science} \bibinfo{volume}{52},
  \bibinfo{pages}{1492--1498}.
%Type = Article
\bibitem[{Wu et~al.(2020)Wu, Mandal and Nguyen}]{Wu2020b}
\bibinfo{author}{Wu, J.Y.}, \bibinfo{author}{Mandal, T.K.},
  \bibinfo{author}{Nguyen, V.P.}, \bibinfo{year}{2020}.
\newblock \bibinfo{title}{{A phase-field regularized cohesive zone model for
  hydrogen assisted cracking}}.
\newblock \bibinfo{journal}{Computer Methods in Applied Mechanics and
  Engineering} \bibinfo{volume}{358}, \bibinfo{pages}{112614}.

\end{thebibliography}
\end{document}